\documentclass[11pt, letterpaper]{article}
\usepackage[margin=1in]{geometry}
\usepackage{amsmath,amssymb,amsthm}
\usepackage{graphicx} 
\usepackage{braket}
\usepackage{dsfont}
\usepackage[allcolors=blue,colorlinks=true,hyperindex=true]{hyperref}
\usepackage{tabularx}
\usepackage{multirow}
\usepackage{nicematrix}
\usepackage[position=top]{subcaption}
\usepackage{booktabs}
\usepackage{siunitx}
\DeclareMathOperator{\TVD}{TVD}

\title{Quantum Image Loading: Hierarchical Learning and Block-Amplitude Encoding}
\author{Hrant Gharibyan, Hovnatan Karapetyan, Tigran Sedrakyan,\\  Pero Subasic, Vincent P. Su, Rudy H. Tanin, Hayk Tepanyan  \\
\\ {\it BlueQubit Inc, 
San Francisco, CA 94105, USA}
\\ {\it Honda Research Institute USA, 
San Jose, California, 95134, USA}}

\begin{document}

\maketitle
\begin{abstract}
  Given the excitement for the potential of quantum computing for machine learning methods, a natural subproblem is how to load classical data into a quantum state. Leveraging insights from~\cite{gharibyan2023hierarchical} where certain qubits play an outsized role in the amplitude encoding, we extend the hierarchical learning framework to encode images into quantum states. We successfully load digits from the MNIST dataset as well as road scenes from the Honda Scenes dataset. Additionally, we consider the use of block amplitude encoding, where different parts of the image are encoded in a tensor product of smaller states. The simulations and overall orchestration of workflows was done on the BlueQubit platform. Finally, we deploy our learned circuits on both IBM and Quantinuum hardware and find that these loading circuits are sufficiently shallow to fit within existing noise rates.
\end{abstract}

\tableofcontents
\clearpage
\section{Introduction}\label{sec:intro}

Quantum computers are racing to fault tolerance with several companies posting roadmaps for achieving that within the next decade. With recent advances in the hardware, a growing number of experiments are believed to be pushing the boundary of what is feasible to simulate classically. However, it is less clear which of the applications of quantum computing, ranging from optimization to simulation to machine learning, is most likely to surpass existing classical competition.

In the domain of machine learning, one difficulty in this comparison to classical methods is the way that classical data is loaded. For example, there are many inequivalent ways to encode an $N$ dimensional vector into the quantum state of $n$ qubits, each with their own challenges of how to practically prepare such a state in physical hardware.

Due to the inherent probabilistic nature of measurement in quantum mechanics, it is natural to consider using quantum states to encode probability distributions. Probability distributions are also important for machine learning since one often works under the assumption that the data in question is from an underlying distribution. Two main approaches for the state preparation task include quantum circuit born machines (QCBMs)~\cite{liu2018differentiable,doi:10.1126/sciadv.aaw9918,coyle2020quantumversusclassicalgenerative} and quantum generative adversarial networks (QGANs)~\cite{Dallaire_Demers_2018,Lloyd_2018,Huang_2021,Zoufal_2019}. Both approaches involve using variational circuits $U(\theta)$ to produce a measurement distribution $q_{\theta}(x) = |\braket{x | U(\theta) | \psi_0}|^{2}$ that mimics a specified probability distribution. Once the variational circuit is learned, it can be loaded many times and measured to produce samples from the distribution $q_{\theta}$. If the distribution is smooth, one can employ techniques to take advantage of the structure with a low entanglement ansatz, as in Ref.~\cite{Iaconis_2024}. In the spirit of ADAPT-VQE~\cite{Grimsley_2019},  Ref.~\cite{shirakawa2021automaticquantumcircuitencoding} proposed loading states by maximizing fidelity with a quantum circuit where the structure and gates are not fixed beforehand. This approach was used to load classical digits in Ref.~\cite{placidi2023mnisq} and for loading stock data in Ref.~\cite{Nakaji_2022}.

In many of these existing applications, numerical simulations were used to get access to the wavefunction. In addition to the exponentially growing memory requirements for simulating a larger numbers of qubits, training of these variational circuits in general suffer from the barren plateau problem~\cite{McClean_2018} (see Ref.~\cite{larocca2024reviewbarrenplateausvariational} for a review). However, in recent work~\cite{gharibyan2023hierarchical}, we were able to encode a multidimensional gaussian distribution on 27 qubits. Variational training on this large quantum state was enabled by a technique called \textit{hierarchical learning} where the structure of the qubit encoding is used to focus on learning fewer qubit circuits before gradually learning the full qubit state.

Here, we extend this technique to loading image data, which can be classically pre-processed to mimic a probability distribution. This is akin to amplitude encoding, though we focus on the resulting classical distribution, meaning we do not necessarily impose specific relative phases in the wavefunction. Since the encoding of the image data also utilizes an ordering on the qubits, we can apply our hierarchical learning techniques. Generically, this encoding will use $\lceil{\log N}\rceil$ qubits, but we also consider the use of block encoding (also explored in~\cite{Huang_2021,Nakaji_2022}), where image data is segmented into smaller images, each of which is encoded independently.

We showcase the effectiveness of our methods by training QCBMs to load images from the MNIST and Honda Scenes dataset. With an aim to load these into existing quantum devices, we optimize our variational circuits with limited circuit depth in mind.

In Section~\ref{sec:QCBMs}, we provide background on QCBMs as well as the mapping between probability distributions and the qubits. In Section~\ref{sec:image-loading}, we discuss how images can be mapped to probability distributions and benchmark the performance of hierarchical loading on various images. We briefly review block amplitude encoding, segmenting the loading of images into separate qubit registers, in Section~\ref{sec:bae}. We proceed to run these image loading circuits on IBM and Quantinuum devices, reporting our results in Section~\ref{sec:hardware-expts} and concluding in Section~\ref{sec:discussion}. Additional details of both the simulation tools and hardware used to run experiments are shared in App.~\ref{sec:software_hardware_details}.

\section{Review of QCBMs and Hierarchical Learning}\label{sec:QCBMs}

Due to the probabilistic nature of measurement in quantum mechanics, measurement outcomes naturally form a probability distribution. Quantum Circuit Born Machines (QCBMs) are quantum circuits whose goal is to prepare a wavefunction that mimics a (classical) target distribution $p(x)$. A variational quantum circuit $U(\theta)$ acting on an initial $n$ qubit state, prepares a trial wavefunction $\psi(\theta)$ that can be expanded in e.g. the computational basis

\begin{equation}
    \ket{\psi(\theta)} = U(\theta) \ket{0^n} = \sum_{b=1}^{2^n} c_\theta(b) \ket{b}  \, .
\end{equation}

Here, the summation over $b$ can be interpreted as the bitstrings of measurement outcomes of the $n$ qubits. The most straightforward mapping is to treat $p(x)$ as a vector whose entries can be indexed with the same discrete labels as the bit string. This is simply the integer representation of the bits, e.g.

\begin{equation}
  \label{eq:binary_encoding}
  b \rightarrow x = \sum_{i=1}^{n} b_{i}2^{n-i} \, ,
\end{equation}
where $x$ is the integer index of the vector. Note that this implies an ordering of the qubits where the first qubit plays the role of the most significant bit.

In this work, we will treat the classical input data to be loaded as a probability distribution. We appropriately normalize the data and keep track of the normalization separately.
The goal of QCBMs is then to tune the parameters $\theta$ such that the generated distribution $q_\theta(x)=|c_\theta(x)|^2$ of the state matches $p(x)$. One can then easily prepare the data in a quantum device if one has a shallow circuit description of $U(\theta)$.

\begin{figure}[t]
  \centering
  \includegraphics[width=.8\linewidth]{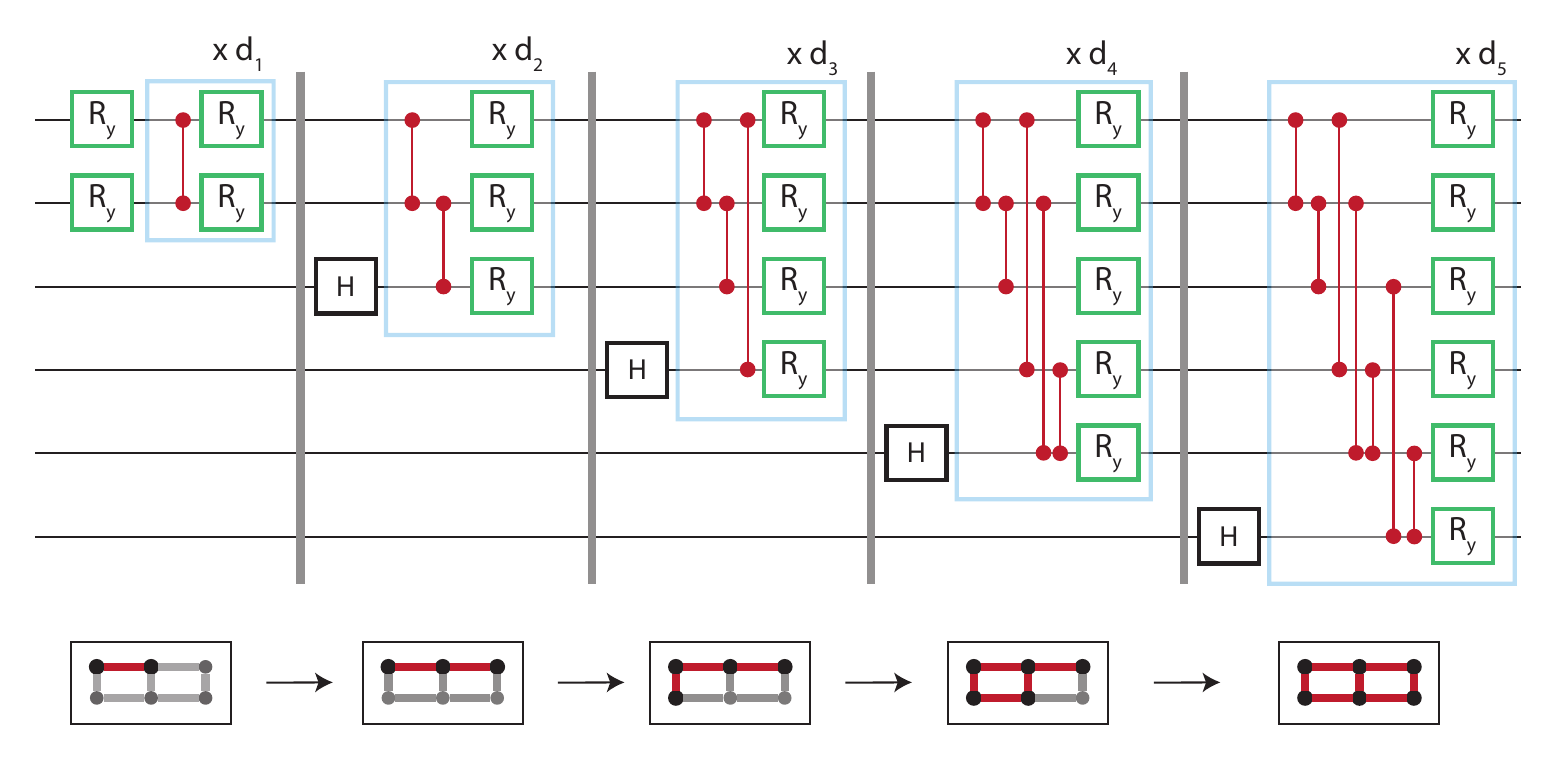}
  \caption{A circuit ansatz with hierarchical architecture on 6 qubits laid out on a grid. Vertical gray lines on the circuit indicate a partial circuit which is variationally learned before using it as input to the next sub-circuit. Starting with two active qubits, we learn variational parameters that approximate the distribution sampled at $2^{2}=4$ data points. Red two-qubit gates correspond to the variational RZZ gates. After training on the two qubit VQC, learned parameters are used as the starting parameters for a 3 qubit VQC with the third qubit initialized in the $\ket{+}$ state and all new parameters are initialized to 0 rather than randomly. The bottom row visualizes the active qubits (black), inactive qubits (gray), and active connectivity (red) for each sub-circuit of the hierarchical VQC.}\label{fig:hierarchical}
\end{figure}

The design and optimization of variational circuits is in general a difficult problem due to vanishing gradients~\cite{McClean_2018,Holmes2022Expressibility,Cerezo_2021} and the general lack of training guarantees.
In our recent article~\cite{gharibyan2023hierarchical}, we have successfully employed a novel hierarchical learning method to load a multi-dimensional Gaussian distribution into a quantum device.
The core principle of hierarchical learning involves a tiered approach to learning distributions, where the process begins by focusing on the most significant digits and progressively integrates additional qubits to represent the less significant digits. This method allows for a structured and prioritized learning sequence that can improve the efficiency and accuracy of the model. The training is done in stages in a layerwise fashion (similarly to~\cite{skolik2021layerwise}), though we further simplify the structure by not including all of the qubits at once, as shown in Fig.~\ref{fig:hierarchical}.

For training of the VQC, we take the loss to be the Kullback-Leibler (KL) divergence between the target distribution $p$ and the generated distribution $q_\theta$.
\begin{equation}
  \label{eq:kl}
  KL(p|q_{\theta}) = \sum_{x} p(x) \log \left(\frac{p(x)}{q_{\theta}(x)}\right)  
\end{equation}

To aid the interpretability of performance, we use total variational distance (TVD) between optimal solution and target data.
\begin{equation}
  \label{eq:tvd}
  \TVD(p,q_{\theta}) = \frac{1}{2} \sum_{x} |p(x) -q_{\theta}(x)| \, .
\end{equation}
This measures the probability mass that is misplaced.

Another useful measure of distance for probability distributions is the classical fidelity, defined as follows
\begin{equation}
  \label{eq:fidelity}
  F(p,q_{\theta}) = \Big( \sum_x \sqrt{p(x) \cdot q_\theta(x)}  \Big)^2 \, .
\end{equation}
This is also the square of the cosine similarity or inner product between the square roots of the distributions. The quantum fidelity $F(\rho, \sigma) = \left(\text{tr} \sqrt{\sqrt{\rho}\sigma \sqrt{\rho}} \right)^{2}$ reduces to the classical fidelity in certain cases, such as when the amplitudes of both states are real and positive.

A state on $n$ qubits measured in the computational basis defines a probability distribution $q_{\theta}(x)$ over $2^{n}$ outcomes. However, if the target distribution has support over a larger domain with $2^{n+1}$ outcomes, one could ``expand'' the distribution $q_{\theta}$ by splitting the probability mass of each outcome into two neighboring points. In terms of the QCBM, this corresponds to adding a least significant qubit in the $\ket{+} = \frac{1}{\sqrt{2}}(\ket{0} + \ket{1})$ state. Similarly, one can coarse grain a distribution by discarding the least significant qubits.

Because we will compare probability mass distributions with different support, arising when comparing QCBMs of different qubit numbers, we introduce the following notation. Let $\TVD_{n}(p, q_{\theta})$ denote the TVD where both distributions are sampled on $2^n$ points. For example, this naturally occurs if $q_{\theta}$ is a QCBM on $n$ qubits. If the target distribution has support on $2^{m}$ points with $m > n$, then $\TVD_{m}(p,q_\theta)$ would be computed with the distribution arising from the addition of $m-n$ qubits in the $\ket{+}$ state. In the case where $m < n$, the distribution is coarse grained by marginalizing over the least significant qubits. This amounts to adding the probability mass of neighboring bins $n-m$ times. We can analogously define $KL_n(p|q_{\theta})$ and $F_n(p,q_{\theta})$.

In our prior work~\cite{gharibyan2023hierarchical}, we developed software packages optimized for the efficient training of QCBMs, leveraging backpropagation and the GPU acceleration capabilities of the BlueQubit infrastructure. Moreover, by integrating the hierarchical learning ansatz, we successfully trained a QCBM on a 27-qubit system, achieving a $4\%$ total variation distance (TVD), significantly surpassing the $15\%$ TVD where naive training strategies plateaued.

\section{Quantum Image Loading}\label{sec:image-loading}

\begin{figure}
    \centering
    \includegraphics[width=0.32\textwidth]{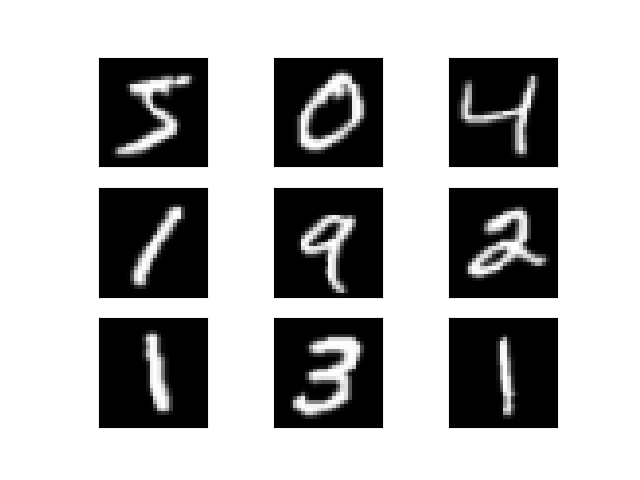}
    \includegraphics[width=0.32\textwidth]{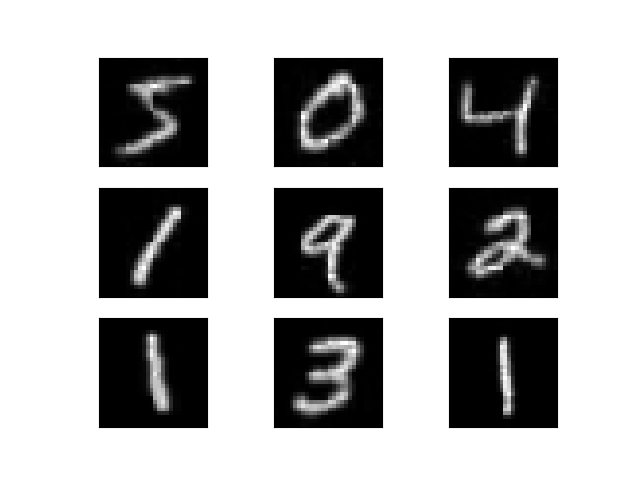}
    \includegraphics[width=0.32\textwidth]{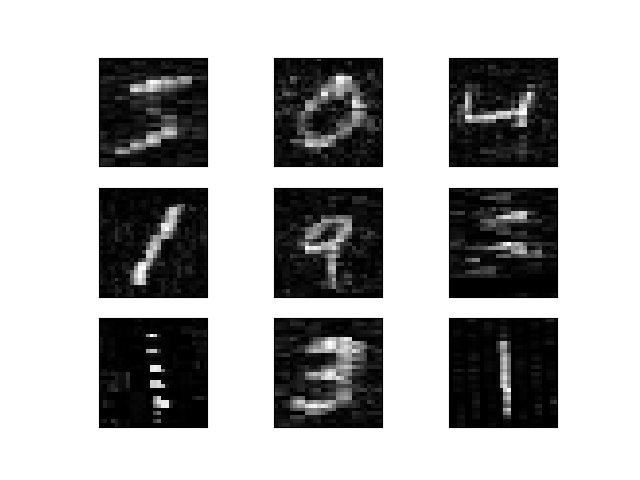}
    \caption{Comparison of encoded digits. The first column contains a sampling of MNIST digits. The middle column contains the loaded distribution using our hierarchical circuits and the right is the same sample of digits loaded by the F80 dataset. Note that our tasks differ slightly in that we encode the image data in the square of the amplitudes while the MNISQ dataset encodes the image data in the absolute value of the wave function amplitudes. These images plot the correspondingly reconstructed images. Our hierarchical circuits were designed to match similar circuit resources to the F80 dataset.}
    \label{fig:q-mnist}
\end{figure}

In this work, we extend our hierarchical learning approach with QCBMs to image data loading. This is a special form of amplitude encoding. Typically, given a data vector $\vec{x}$, amplitude encoding refers to the preparation of a wavefunction $\ket{\psi} = \sum_i \tilde{x}_i \ket{i}$ where $\tilde{x}_i = \frac{x_i}{|\vec{x}|}$. However, we take a slightly different approach by clasically preprocessing the data to be a probability vector (e.g. $\sum_i p_i = 1$ and $0 \leq p_i \leq 1$). Preparing a QCBM for this distribution is then akin to doing amplitude encoding with entries $\sqrt{p_i}$.

We apply this technique to image data coming from MNIST and the Honda Images dataset~\cite{honda-data-set}. In both cases, the gray-scale pixel intensity is rescaled to fit between 0 and 1. The pixel data is then treated as a vector and normalized to be a valid probability distribution. If correctly prepared, the histogram of measurement outcomes should yield the normalized image.

Just as in the case of loading probability distributions, the mapping between the wavefunction basis elements and pixel locations dictates a structure on the qubits. Suppose there are $n$ qubits. We can label them as $q_{v_1},\cdots, q_{v_j}, q_{h_1},\cdots,q_{h_k}$ where $j + k = n$. There are most and least significant qubits for both dimensions of the 2D image data. Thus, in our hierarchical loading where qubits are gradually introduced, we start with the most significant ones $q_{v_1}, q_{h_1}$. This is akin to first learning a coarse grained image. The upshot we see is that we can achieve higher fidelities with a fixed amount of resources.

It is worth noting that our approach to image loading here differs somewhat from the standard amplitude encoding (up to a squareroot). However, the practical difference for machine learning applications remains unknown. When thinking of amplitude encoding as a feature encoding in Hilbert space, it simply corresponds to one type of kernel, whereas our procedure would lead to a modified kernel. Which one works better may depend on the application. In parallel work~\cite{honda2}, we explore the use of the typical approximate amplitude approach to image loading for quantum machine learning for classification.

\subsection{MNIST}
The MNIST dataset~\cite{deng2012mnist} has played an influential role in classical machine learning, serving as the first test ground for testing different methods. Here, we turn our attention to loading this classical data into a quantum register.
MNIST images consist of 28x28 pixel grayscale images, which we subsequently normalize to the range 0 to 1 and enforce the probability constraint. The grid is transformed into a 784 dimensional vector, requiring a mere 10 qubits.

The variational circuit we use assumes a qubit connectivity of nearest neighbors on a 2x5 grid, similar to that shown in Fig.~\ref{fig:hierarchical}. The top row qubits map to the horizontal positions while the bottom row represents the vertical direction. From left to right we have most to least significant qubits. Our hierarchical approach starts with the leftmost 2 qubits and gradually adds 2 at a time. At each stage, only the part of the variational circuit which involves active qubits are trained using gradient techniques. When including more active qubits, their parameters are initialized to 0. The only randomization occurs with the first layer of the circuit.

Previously, a library of quantum circuits preparing these images was released~\cite{placidi2023mnisq}. The circuits were generated with a procedure called Automatic Quantum Circuit Encoding (AQCE)~\cite{shirakawa2021automaticquantumcircuitencoding} where gates and their locations are placed to maximize the quantum fidelity with the target state. This procedure stops after a threshold of fidelity or a maximum circuit depth. The smallest circuits they release are part of the F80\footnote{Note that the fidelity they consider is $|\braket{\psi|\phi}|$ as opposed to the square.} dataset, where the threshold fidelity is $80\%$ and the maximum number of 2-qubit gates is 65.

Applying our hierarchical methods, we load the same images with a pre-determined circuit structure for all digits and a fixed 2-qubit gate count of 65 to match. Quite generically, we achieve a larger fidelity despite a more rigid circuit ansatz. For a visual comparison of the loaded images, see Fig.~\ref{fig:q-mnist}.

\subsection{Honda Scenes Dataset}\label{sec:hsd-dataset}
To test our data loading technique on more complicated images, we draw from the Honda Scenes dataset~\cite{honda-data-set}. The Honda Scenes is an extensive, labeled collection designed for dynamic scene classification. It encompasses 80 hours of varied, high-quality driving footage from the San Francisco Bay area. This dataset features time-based annotations detailing various aspects such as road locations, surrounding environment, weather conditions, and the state of the road surface.

The Honda dataset features 1024x2048 images as frames\footnote{Original images were 1080x1920, but we reshaped them for convenience.}. Compared to MNIST images, these require 21 qubits to represent the amplitudes of the entire image. We deal with the gray-scale for simplicity, but the adaptation to color images can be straightforwardly accomplished by treating the three color channels separately.

\begin{figure}
    \centering
    \includegraphics[width=0.9 \textwidth]{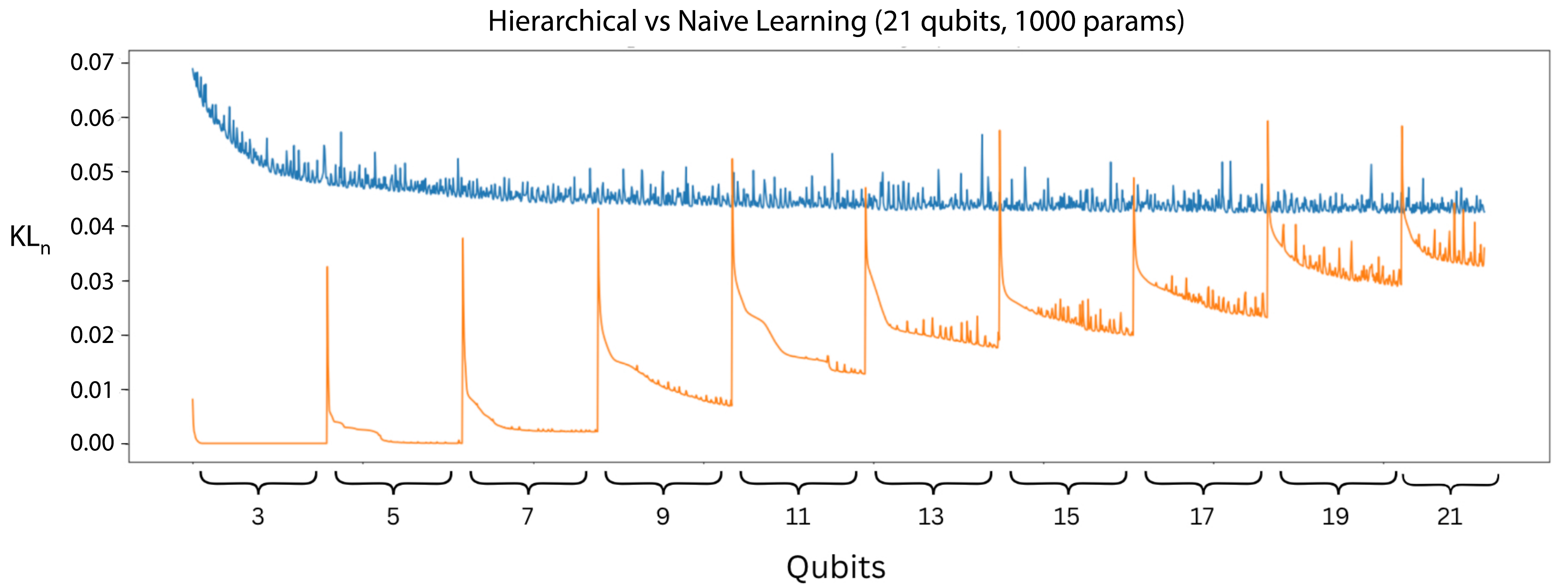}
    \caption{Training of 21 qubit amplitude loading quantum circuit with QCBMs. The blue is training without hierarchical learning, that achieves $KL_{21}$ of 0.042 in the end of training. The yellow line represents the results of training using a hierarchical method, in which qubits are added iteratively. This approach resulted in achieving a KL divergence ($KL_{21}$) of 0.034 by the end of the training period. The yellow line measure $KL_n$ from 3 to 21 qubits, that's the course grained KL divergence for each qubit choice.}
    \label{fig:q-gpu-training-kl}
\end{figure}

\begin{figure}[ht]
    \centering
    \includegraphics[width=0.55\textwidth]{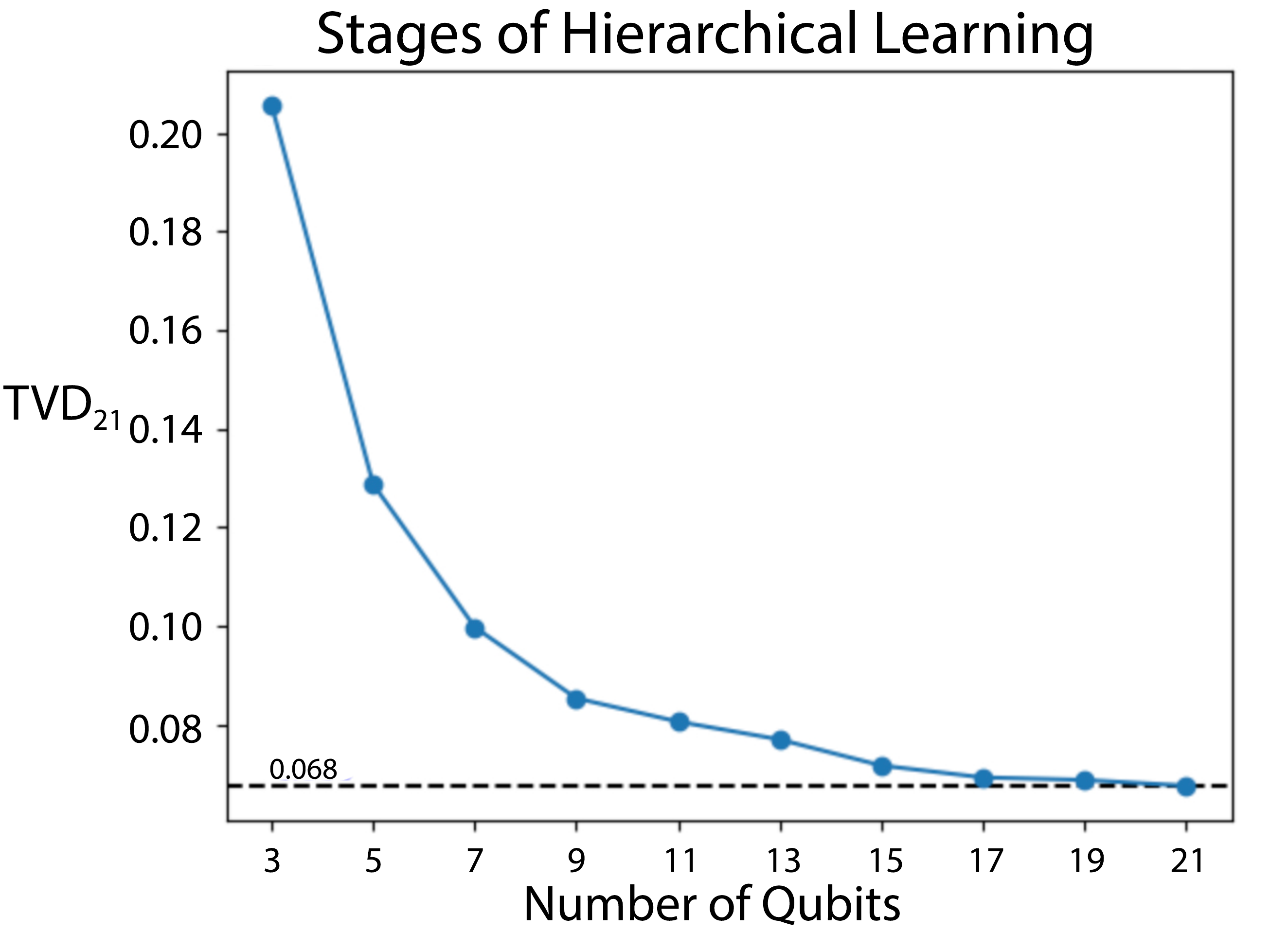}
    \caption{Fine-grained total variational distance ($\TVD_{21}$) for the hierarchical learning strategy, as the circuit expands from 3 qubits to 21 qubits, resulted in a final TVD of $6.8\%$. }
    \label{fig:q-tvd21}
\end{figure}

For the training, we again minimize the KL divergence between the target distribution and the classical distribution obtained from measurements of the prepared wavefunction. To highlight the advantage that hierarchical training provides, we compare the performance of the hierarchical circuit with a variational circuit that starts with the full connectivity of the 2D grid. The number of variational gate parameters was kept fixed at 1100. In Figure~\ref{fig:q-gpu-training-kl}, we report $KL_{n}$, which depends on the number of active qubits, for the hierarchical circuit and the full $KL_{21}$ for the non-hierarchical circuit.
The hierarchical strategy attained a KL divergence of $0.034$ at the end of training, in contrast to the straightforward approach, which plateaued at a KL divergence of $0.042$.

\begin{figure}
    \centering
    \includegraphics[width=0.45\textwidth]{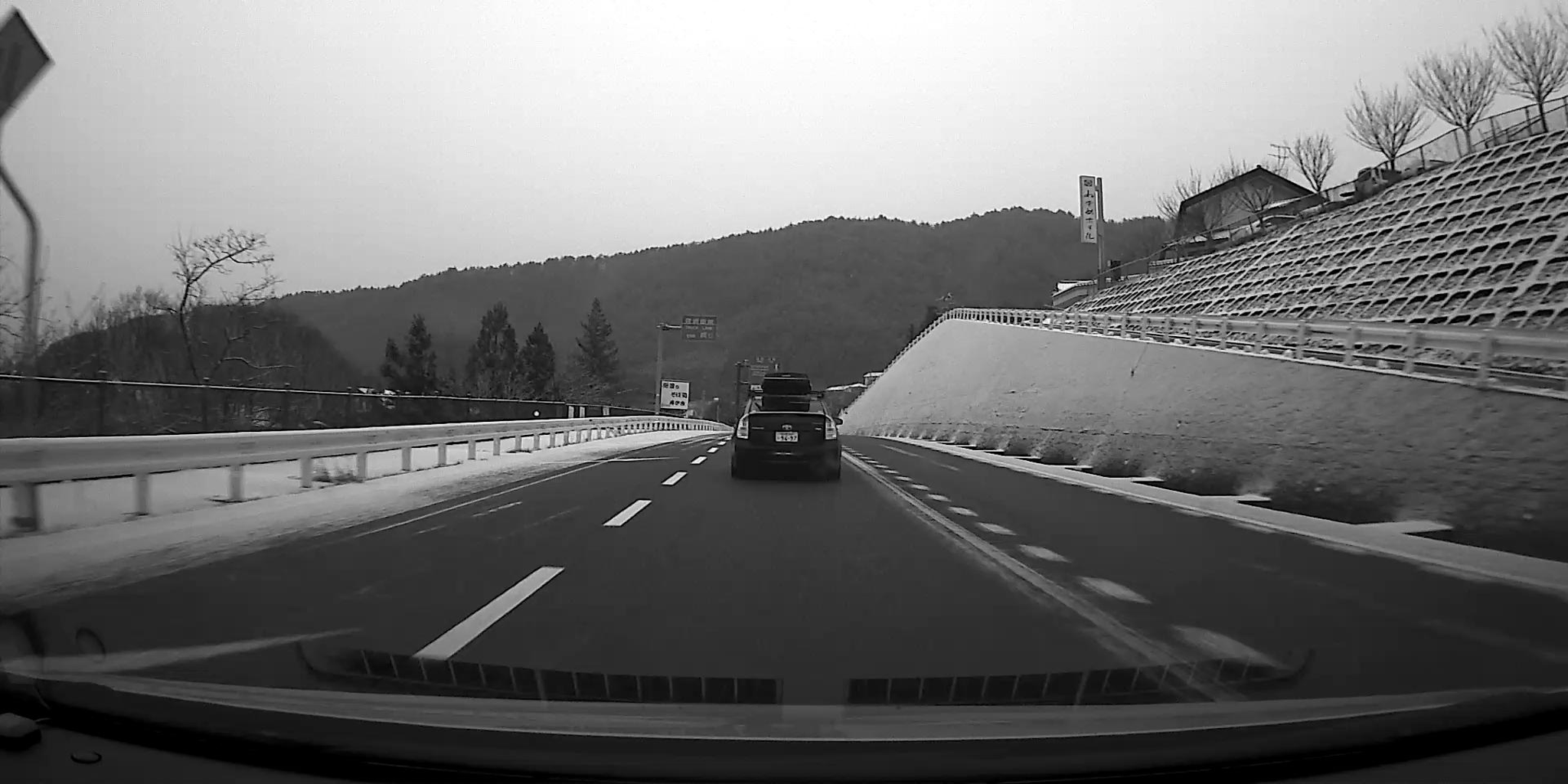}
    \includegraphics[width=0.45\textwidth]{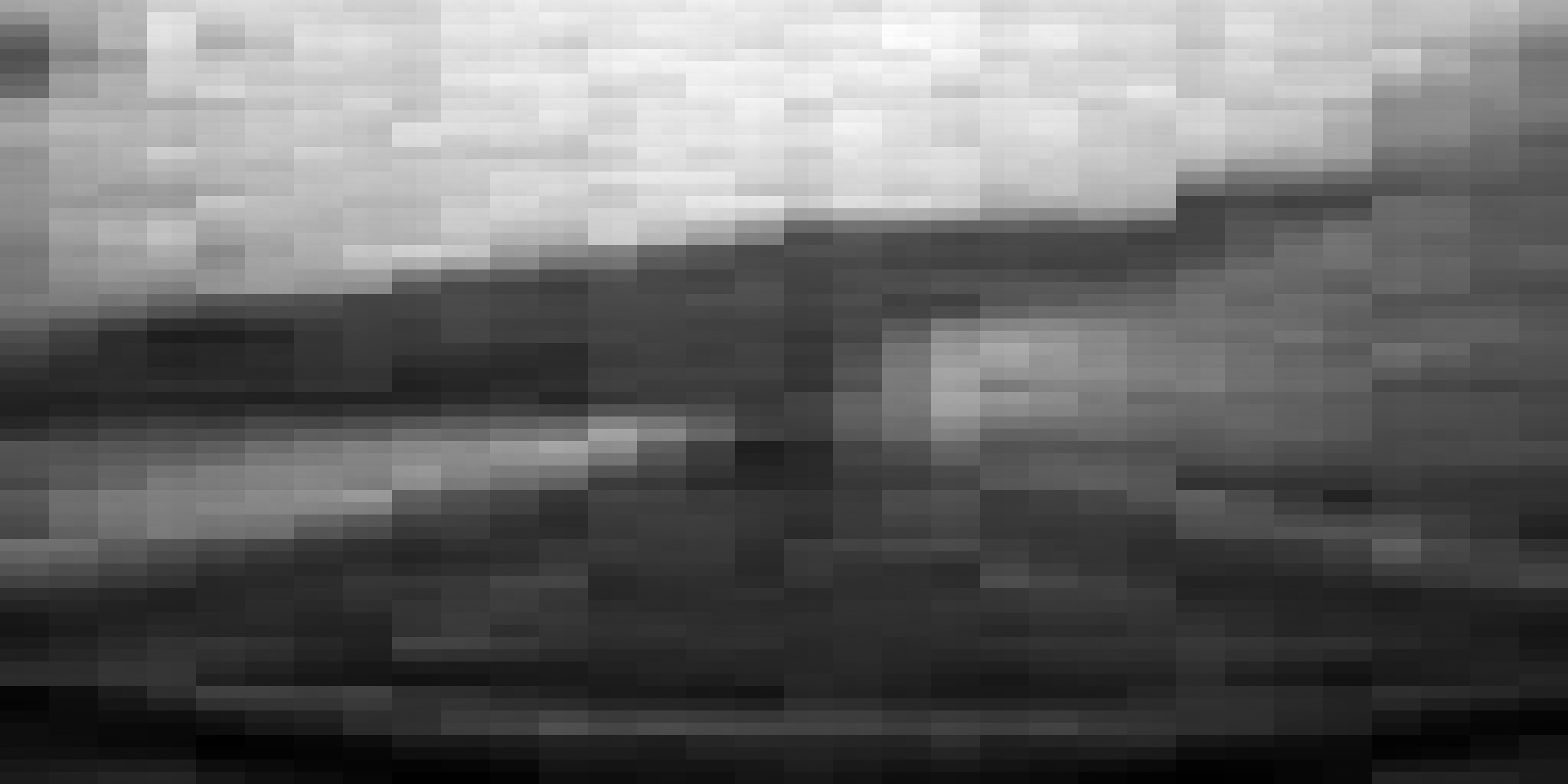}
    \caption{{\it Left: } Sample image (greyscaled and resized to 1024x2048 dimension) from Honda's dataset. {\it Right} Amplitude loaded version of the image into a 21 qubit quantum state with BlueQubit's QCBM and hierarhical learning method ($6.8\%$ TVD)}
    \label{fig:q-honda-data}
\end{figure}

To better understand the intermediate training progress for the hierarchical approach,
we present in Figure~\ref{fig:q-tvd21} the \(\TVD_{21}\), e.g. the TVD at the full resolution even when only a fraction of the qubits are active. While the KL divergence values seem to get worse as more qubits are introduced, that is due in part because $KL_{n}$ probes the image at finer and finer scales.
Notably, as we increase the number of qubits, there's a significant improvement in accuracy, culminating in a \(\TVD_{21}\) of \(6.8\%\).

The final image reconstructed from this training is shown in Figure~\ref{fig:q-honda-data}. It's important to highlight that despite the relatively low \(\TVD\), the quality of the reconstructed image is not particularly high; it primarily captures broad features such as the sky and road, but lacks detail. There's potential for further improvement with additional training, though the challenge escalates given the extreme compression of 2 million pixels into a 21-qubit, 1100-parameter quantum circuit. In the following section, we introduce an approach to image data loading that enables more effective image loading schemes.

\section{Block-Amplitude Encoding}\label{sec:bae}

\begin{figure}
    \centering
    \includegraphics[width=0.32\textwidth]{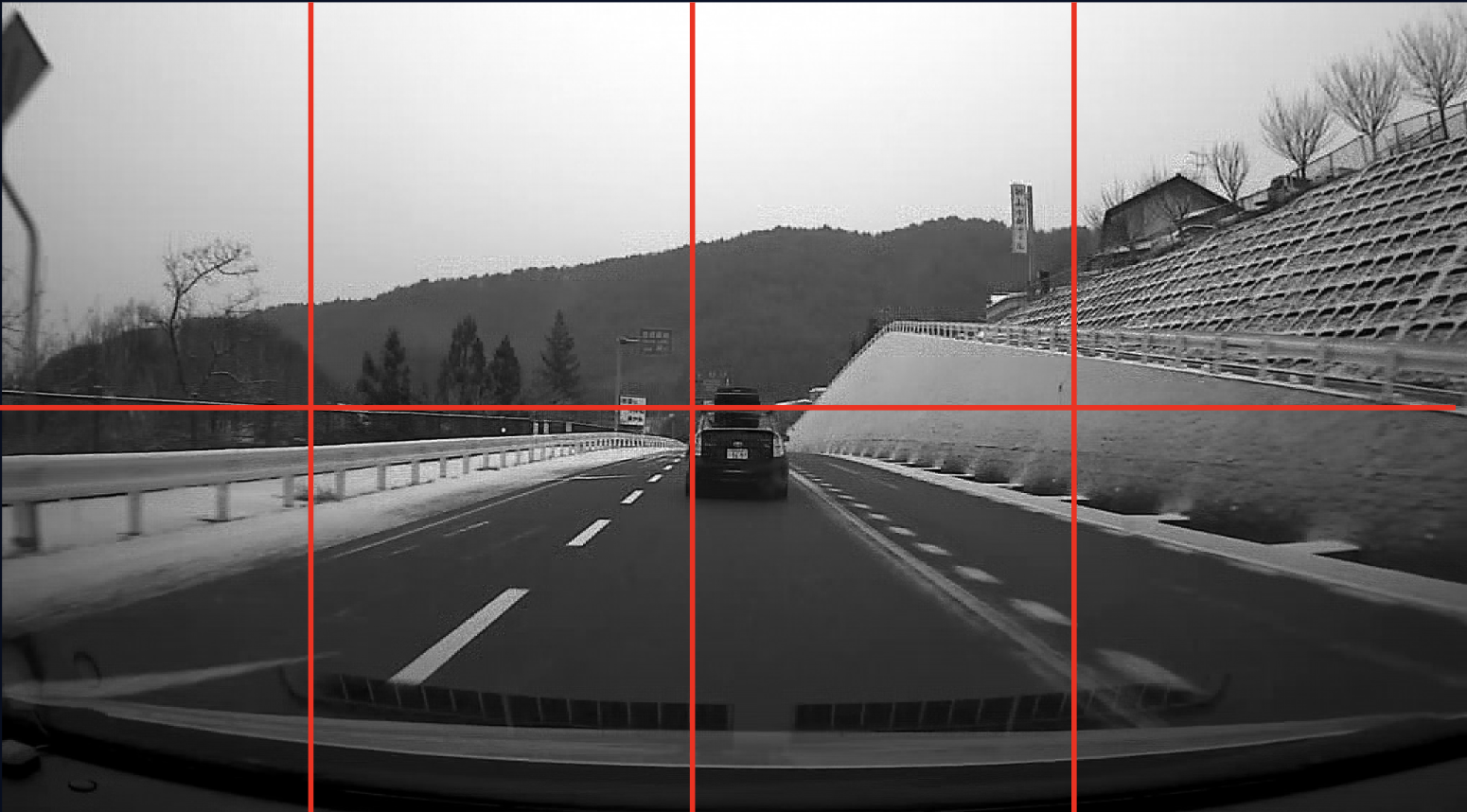}
    \includegraphics[width=0.32\textwidth]{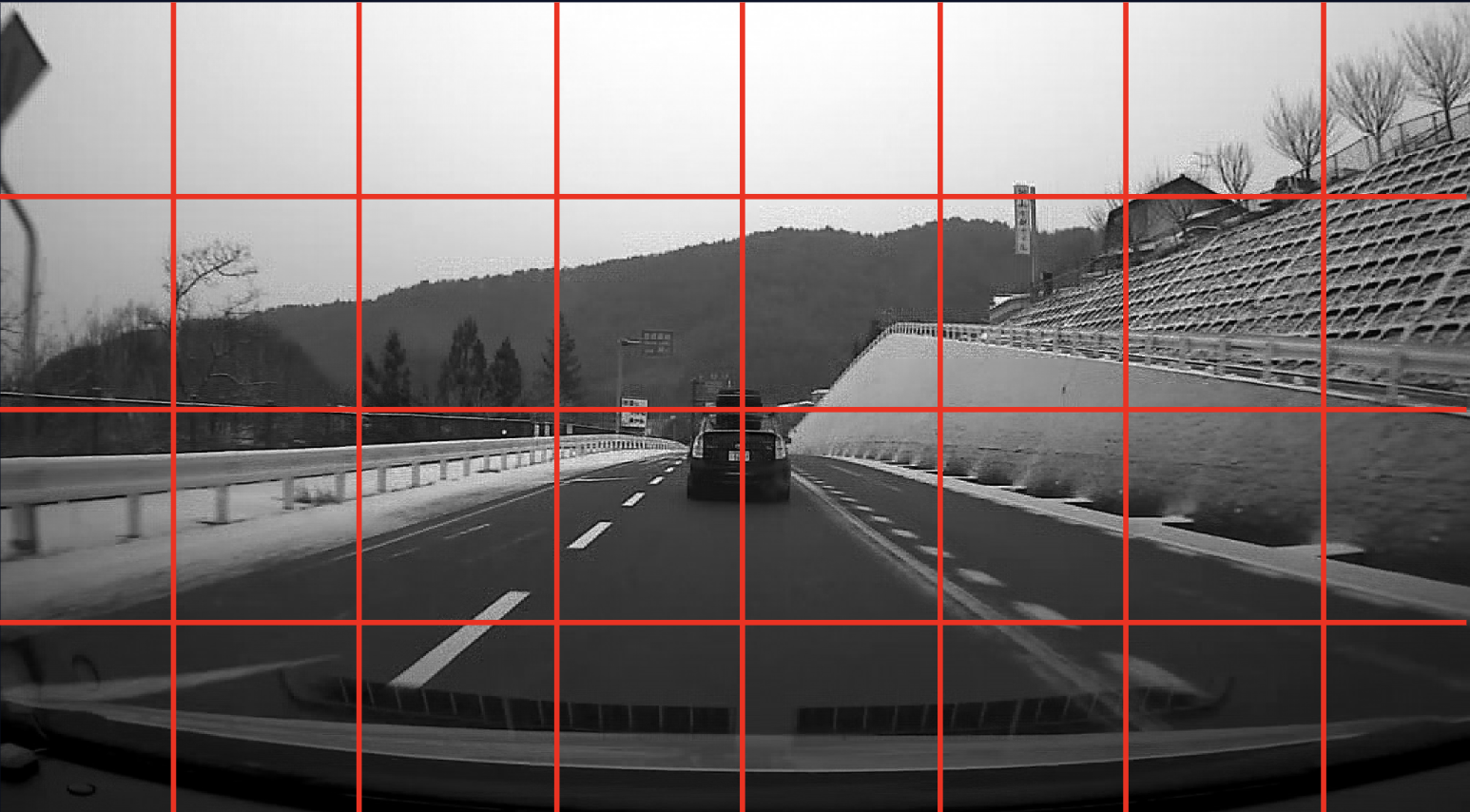}
    \includegraphics[width=0.32\textwidth]{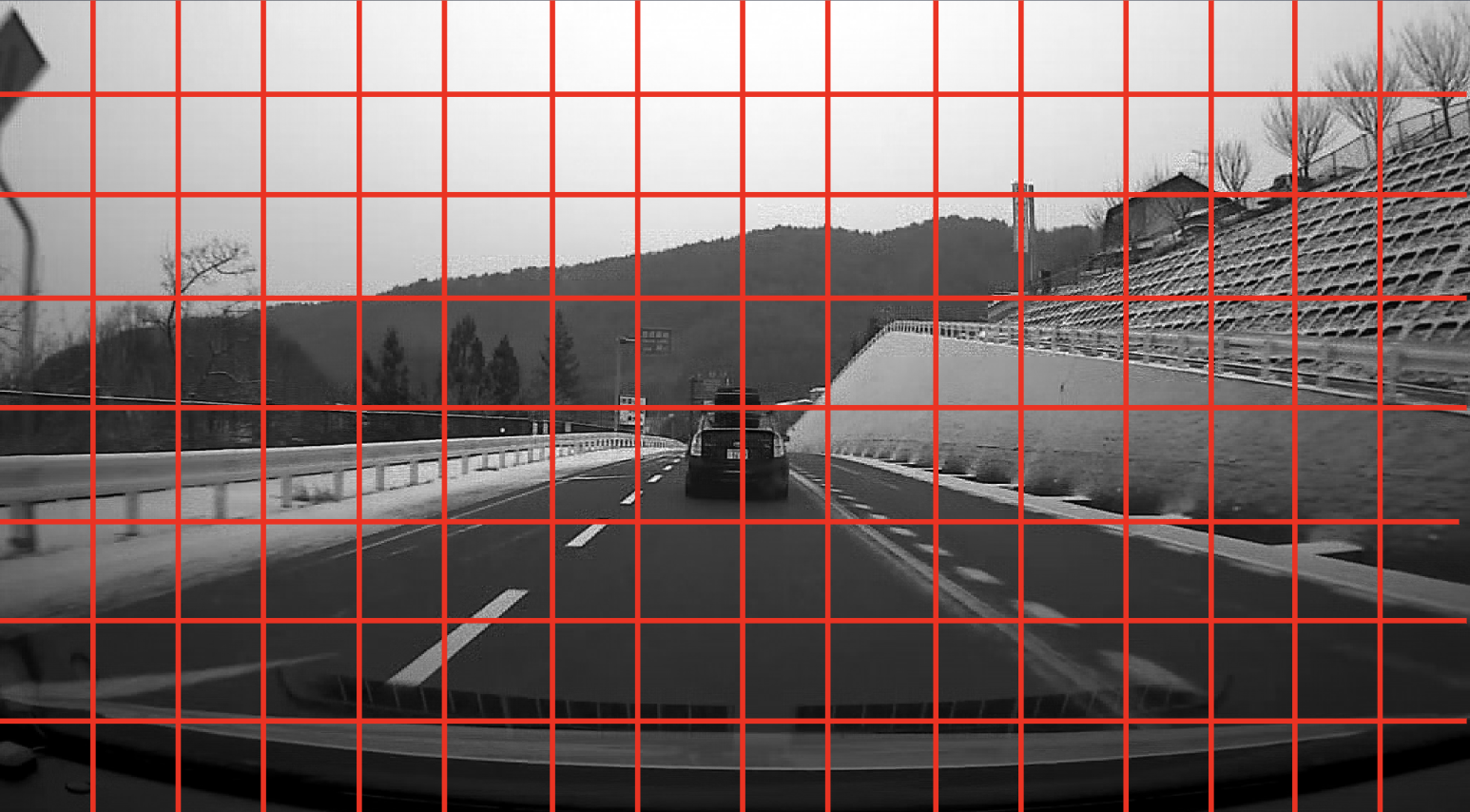}
    \caption{{\it Left: } Original image is divided into 8 blocks. {\it Center} Original image is divided into 32 blocks.  {\it Right} Original image is divided into 128 blocks.  }
    \label{fig:q-image-blocks}
\end{figure}

Here we introduce an encoding technique termed Block-Amplitude Encoding (BAE), which fully capitalizes on the capabilities of QCBMs and the hierarchical learning methods outlined in preceding discussions. BAE is designed to achieve significantly precise image approximations by partitioning the image into discrete blocks and individually mapping these blocks onto qubits. This method allows for a granular level of control and flexibility in image representation within a quantum computing framework. For the original image size of $d_1 \times d_2$, the height of the image (\(d_1\)) is divided by the parameter \(b\), and the width (\(d_2\)) is divided by \(2b\) to facilitate the creation of square blocks. Consequently, the total number of blocks is \(2b^2\). In Figure~\ref{fig:q-image-blocks}, we illustrate the image segmented into blocks, with sizes ranging from 8 to 128 blocks.

The image is first partitioned into square blocks, after which the blocks are amplitude-encoded onto quantum circuits. Each of the circuits corresponding to separate blocks is not entangled with the rest, thus reducing the number of entangling two-qubit gates required to load the whole image. Due to the modularity of this procedure, blocks can be loaded using hierarchical learning, by training each of the circuits separately and composing them together to allow for the loading of the complete image. With BAE, it is possible to fine-tune the block size, in order to achieve an optimal trade-off between the loaded image quality and the requirement for quantum resources.

This approach represents an intermediate strategy between angle encoding, where each pixel is mapped to a single qubit (using $O(d_1 \times d_2)$ parameters) and full amplitude encoding (that requires $O(\log (d_1 \times d_2))$ parameters), providing a unique flexibility to navigate the spectrum between these two extremes through the selective determination of block numbers. More specifically, for the block parameter \(b\), the total number of qubits required to load the entire image is given by
\begin{equation}
    O\left(b^2 \log \frac{d_1 \times d_2}{b^2}\right)
\end{equation}
where each block is encoded into \(\log \frac{d_1 \times d_2}{b^2}\) qubits. Note that we keep track of the normalization of each block separately.

\begin{figure}
    \centering
    \includegraphics[width=0.3\textwidth]{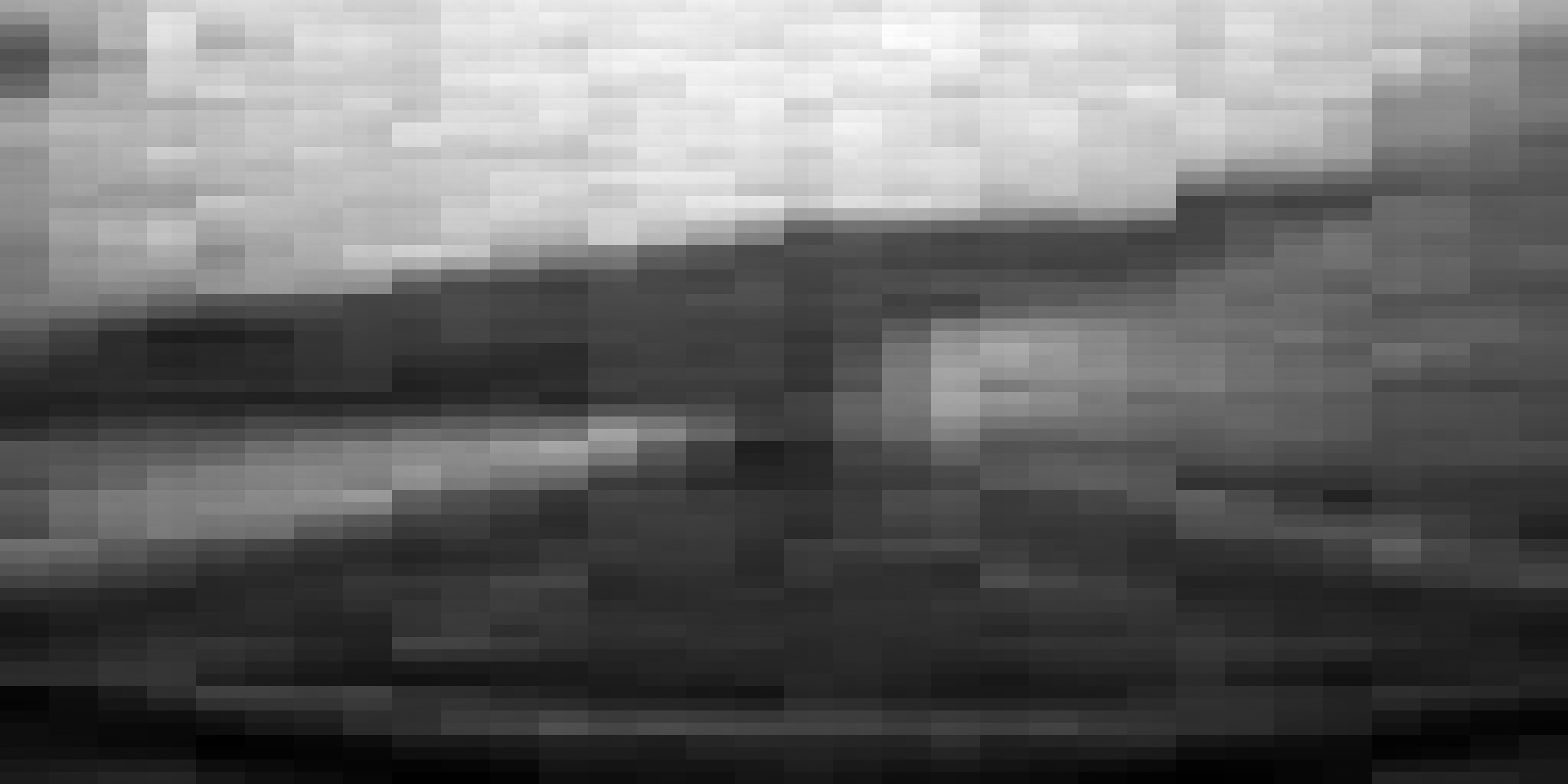}
    \includegraphics[width=0.3\textwidth]{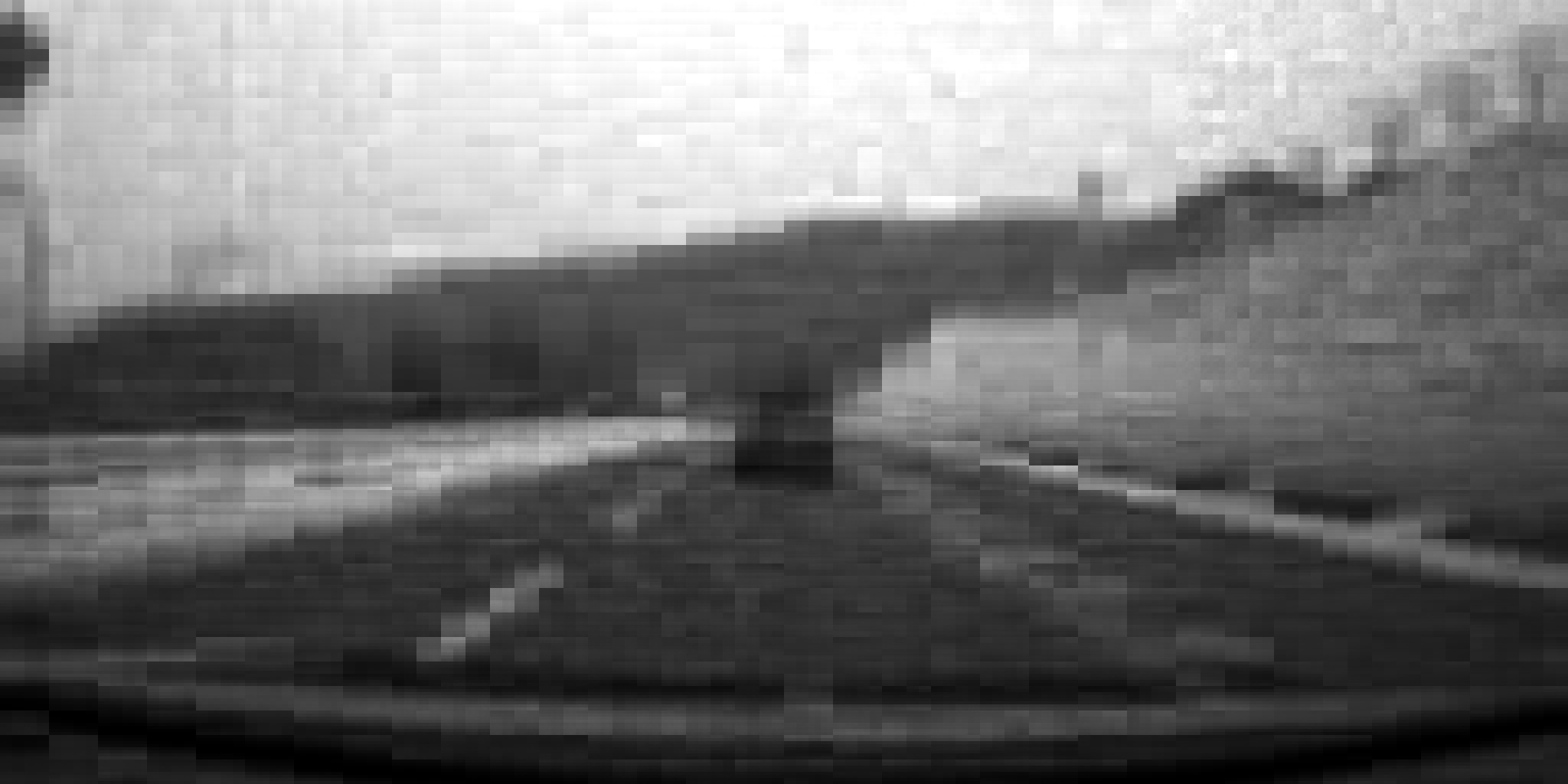}
    \includegraphics[width=0.3\textwidth]{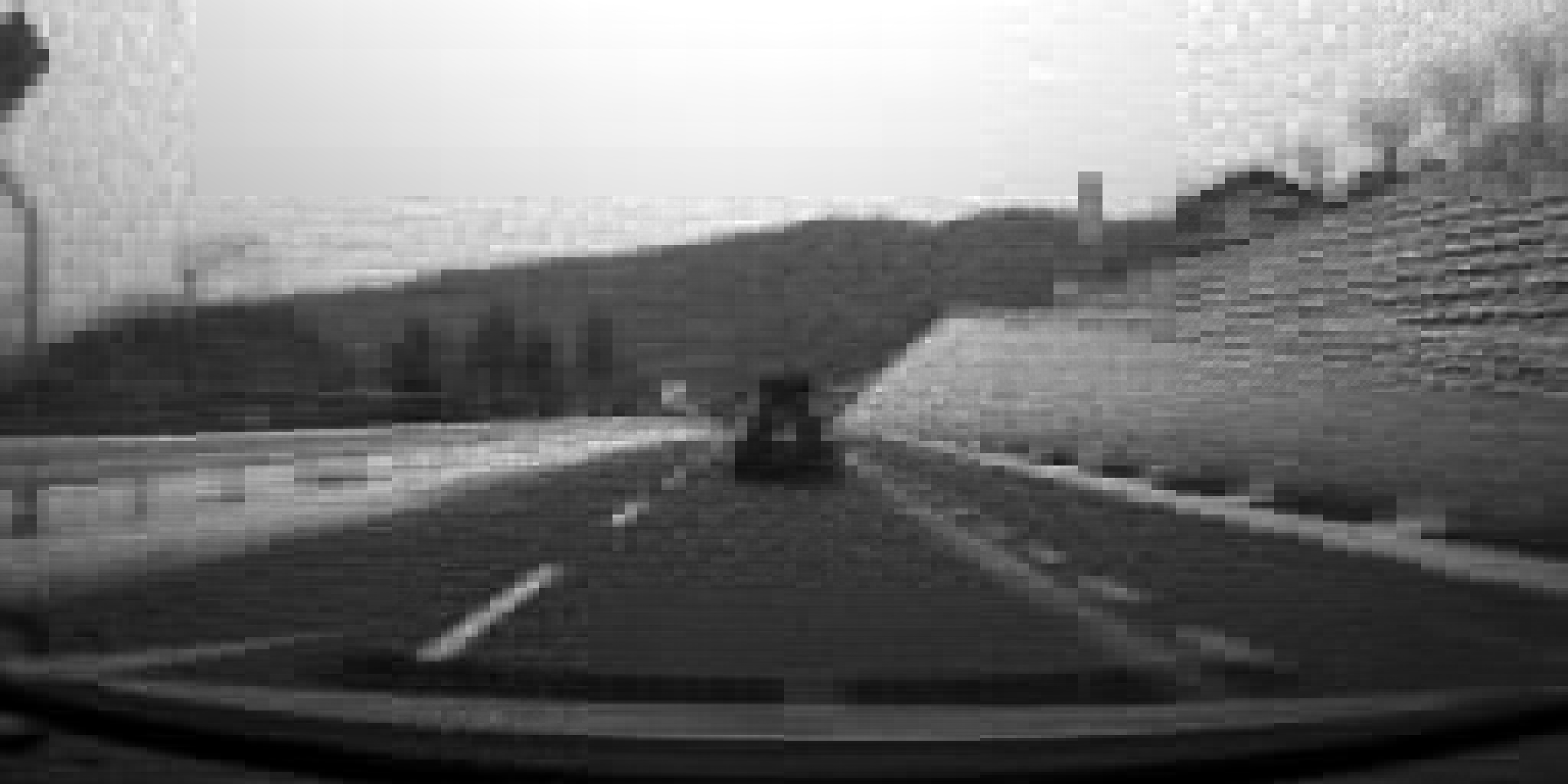}
    \includegraphics[width=0.3\textwidth]{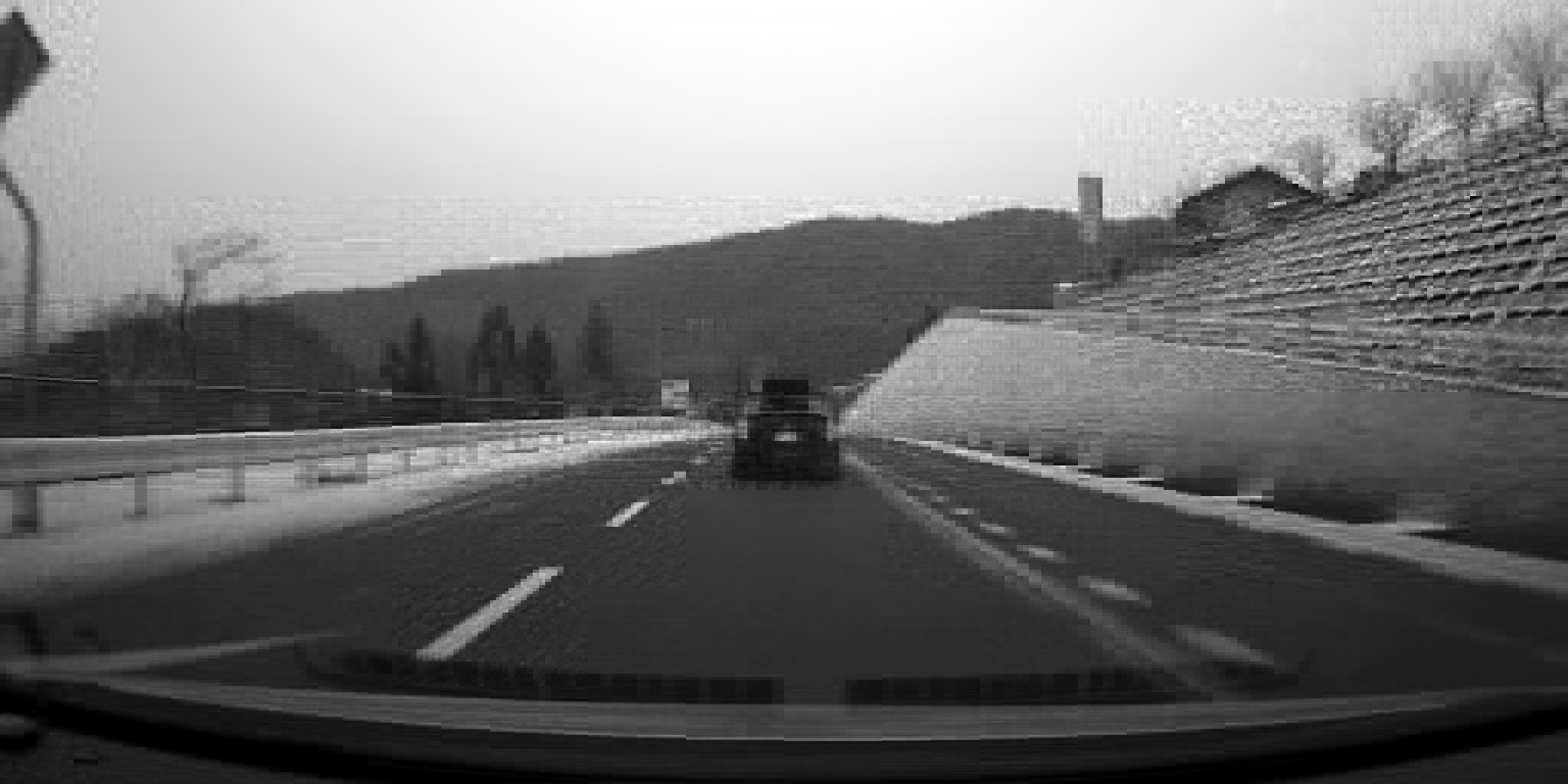}
    \includegraphics[width=0.3\textwidth]{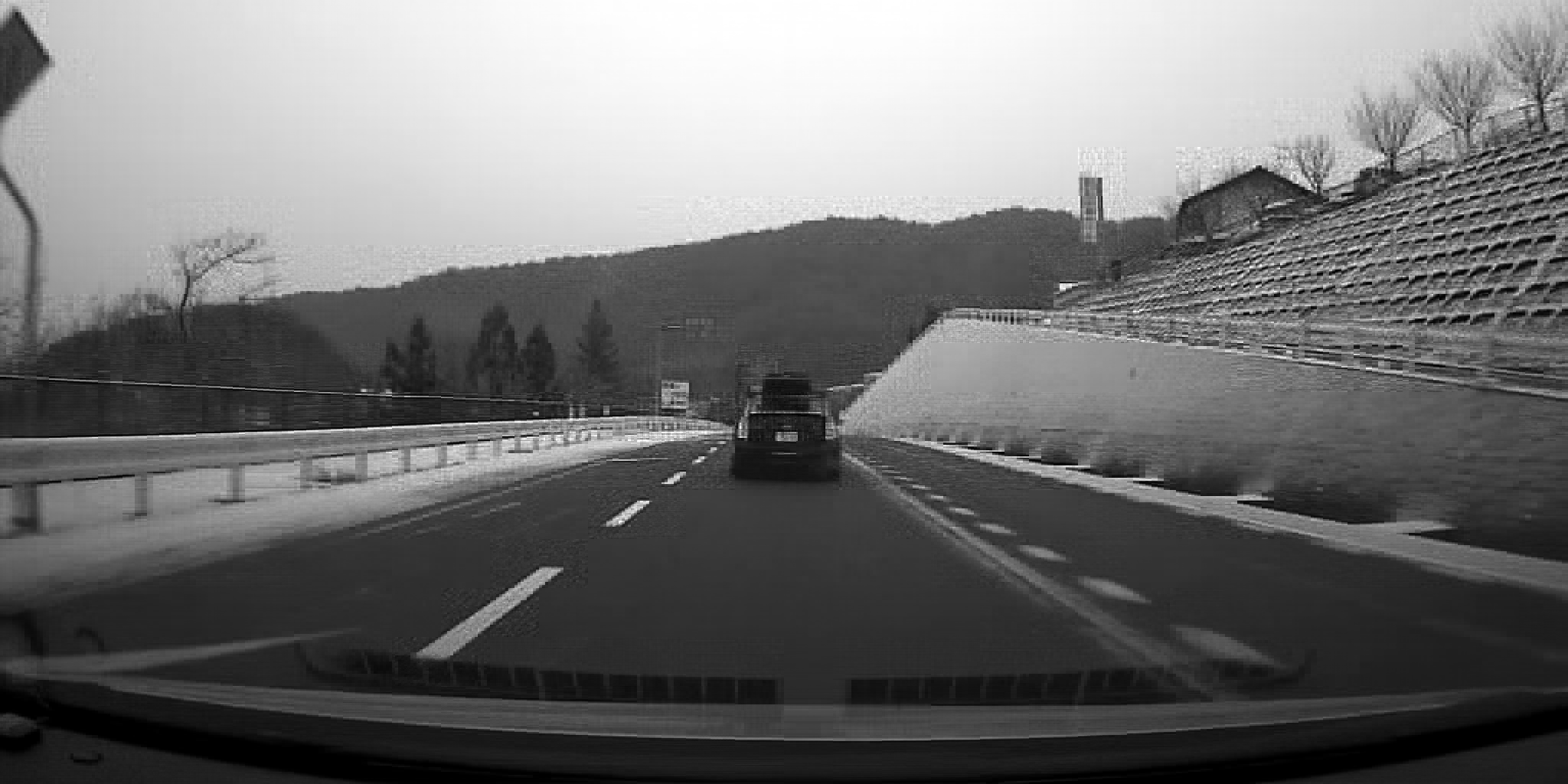}
    \includegraphics[width=0.3\textwidth]{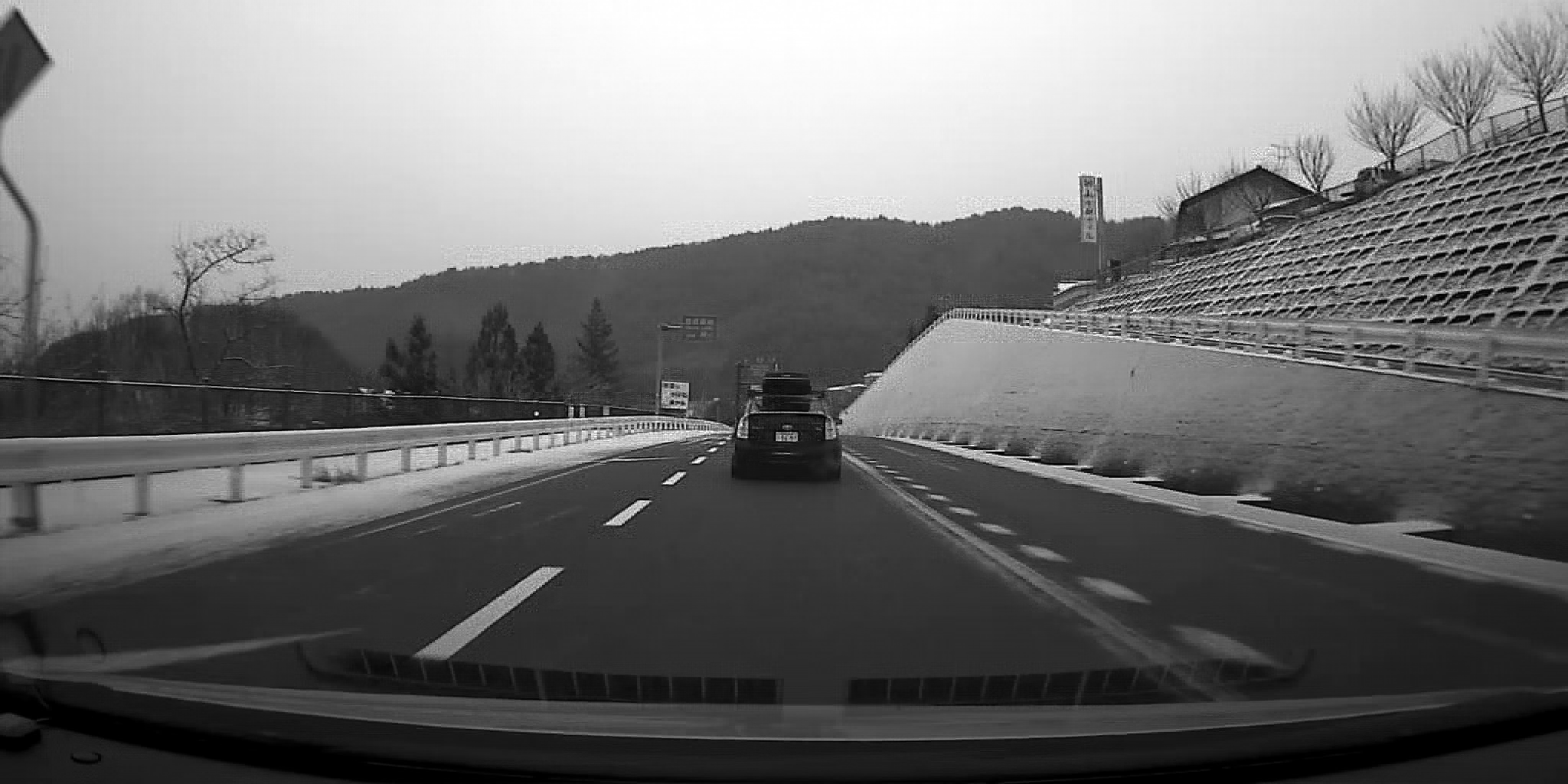}
    \caption{From the top left to bottom right images: 1 block, 21 qubit; 8 blocks, $8 \times 18=144$ qubits; 32 blocks, $32\times16 = 512$ qubits; 128 blocks, $128\times 14 = 1792$ qubits; 512 blocks, $512 \times 12 = 6144$ qubits; 2048 blocks, $2048 \times 10 = 20K$ qubits}
    \label{fig:q-images-bvals}
\end{figure}

For a visual comparison of the image loading quality, refer to Figure~\ref{fig:q-images-bvals}, where we display images ranging from single block loading to examples with 2048 blocks (each block is loaded on 10 qubits), based on the initial image size of 1024x2048. In Table~\ref{tab:bae-numbers}, we detail the outcomes of BAE experiments, spanning from a single block to 2048 blocks, including specifics on the number of qubits per block, total qubits, overall circuit parameters, and TVD. It is noteworthy that, at approximately 30 blocks, the images accumulate sufficient detail and features to consider the development of quantum classifiers based on those quantum loaded circuits.

\begin{table}[ht]
\centering
\begin{tabular}{|c|c|c|c|c|c|}
\hline
{\bf \# Blocks} & {\bf Qubits per Block} & {\bf Total \# Qubits} & {\bf Total \# parameters} & {\bf TVD} \\
\hline
1 & 21 & 21 & 1100 & 6.8\% \\
\hline
8 & 18 & 144 & 3328 & 4.9\% \\
\hline
32 & 16 & 512 & 10K & 3.9\% \\
\hline
128 & 14 & 1792 & 33K & 3.1\% \\
\hline
512 & 12 & 6144 & 96K & 2.4\% \\
\hline
2048 & 10 & 20K & 270K & 1.6\% \\
\hline
\end{tabular}
\caption{Results of the with BAE experiments ranging from single block to 2048 blocks, detailing the number of qubits per block, total qubits, total circuit parameters, and TVD.}
\label{tab:bae-numbers}
\end{table}

\section{Quantum Hardware Experiments on IBM and Quantinuum}\label{sec:hardware-expts}

Both hierarchical learning and block amplitude encoding are strategies for loading data in quantum registers in a hardware friendly manner. In particular, BAE allows each block to be loaded in parallel. This has the additional advantage that individual blocks can be simulable even when the full quantum data may not fit in a classical simulator. One could, for example, potentially find useful applications of building a quantum classifier on this data that would be difficult to simulate classically.

To that end, we deploy our hierarchical circuits on existing quantum hardware to assess their feasibility. We perform both experiments encoding the full 21 qubit state with no block encoding and a separate experiment with 72 qubits where the image is first downsampled and then partitioned into 6 blocks of 12 qubits each. In principle, one could verify that the distribution was loaded properly by sampling bit strings, yielding a histogram mimicking the normalized image data. However, to get sufficient statistics on a distribution over $2^{21}$ points, one needs quite a larger number of measurements. Additionally, due to measurement and readout errors that affect all of the qubits, we instead probe more coarse grained statistics of the data as a sanity check, namely single Pauli Z expectation values. 

For a single qubit, written as $\ket{\phi} = \alpha \ket{0} + \beta \ket{1}$, the $Z$ expectation value is just given as $\braket{Z} = |\alpha|^2 - |\beta|^2$. Relatedly, one can ask about the probability of measuring a $0$, given just by $|\alpha|^2$. Due to the normalization constraint $|\alpha|^2 + |\beta|^2 = 1$, these essentially yield the same information, but the positivity of the latter helps in plotting. 

Since we encode the structure of the image in the binary mapping of qubits, the marginal state on each qubit gives information about the intensities on half of the image. For the most significant qubit $q_{v_1}$, $P_{q_{v_1}}(0)$ measures the sum of intensities on the top half. For $q_{v_2}$,  $P_{q_{v_2}}(0)$ corresponds to the sum of intensities on the first and third quarters, making up half of the image. To avoid unnecessary clutter, we will drop the parenthesis and implicitly take $P_i$ to be the probability of measuring the $i$-th qubit in the 0 state.

Note that this issue of noisy measurement is a bottleneck on the classical verification and not for the purposes of building, for example, a quantum classifier on top of the quantum data. In~\cite{honda2}, we indeed explore this avenue where the result of the classification is encoded in at most a few qubits rather than the joint distribution of several qubits.

\subsection{Quantinuum}\label{ssec:quantinuum}
Quantinuum features a trapped ion architecture with any-to-any interactions owing to its flexible positioning of atoms to their interaction zone. We perform image loading based on hierarchical circuits from Figure \ref{fig:hierarchical} with gates adapted to what is available on the hardware, specifically RY and RZ gates for the rotational layers and RZZ gates for the entangling layers. The optimization of circuits is done classically as explained in Section~\ref{sec:software_hardware_details}.
We ran 3 experiments on the 20 qubit Quantinuum H1 device loading a 1024x1024 image of a dry road from the HSD dataset from Section \ref{sec:hsd-dataset}.
We run 3 experiments with increasingly more gates to test the hardware performance in various regimes. The 3 circuits are described in the Table~\ref{tab:quantinuum-exp-table}.

\begin{table}[ht]
\centering
\begin{tabular}{|c|c|c|c|c|c|c|c|}
\hline
{\bf Name} & {\bf \# Qubits} & {\bf \# 2Q Gates} & {\bf Gate Count} & {\bf TVD$_{20}$} & {\bf \# Shots} & {\bf $\sum_i|P_i - P^{*}_i|$} \\
\hline
Small  & 20 & 149 & 430  & 6.8\% & 200 & 0.084 \\
\hline                            
Medium & 20 & 195 & 550  & 6.4\% & 100 & 0.205 \\
\hline                            
Large  & 20 & 475 & 1230 & 5.9\% & 100 & 0.277 \\

\hline
\end{tabular}
\caption{Details of the 3 circuits run on Quantinuum H1 device. TVD$_{20}$ is measured for the simulated wavefunction. For the hardware, we quantify the results by seeing how the measured probabilities $P_i$ differ from the true probabilities $P^{*}_i$ is a metric to quantify the quality of quantum results.}
\label{tab:quantinuum-exp-table}
\end{table}

Generally, the deeper circuits allow us to find a better approximation of the image, as measured by the TVD from the ideal circuit. With relatively few shots, however, one cannot verify the probability distribution over a million pixels.
Instead, we benchmark the coarse grained statistics of the most significant qubits, namely their probability to be found in the 0 state, $P_i$, compared to the true values $P^{*}_i$. The interpretation of these correspond to the fraction of intensity in the half of the image that the qubit toggles between as discussed in Sec.~\ref{sec:image-loading}.

\begin{figure}[ht]
    \centering
    \subfloat[]{\includegraphics[width=0.45\textwidth]{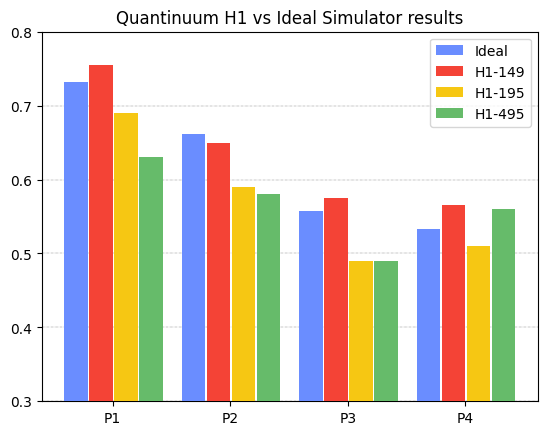}
    \label{fig:quantinuum-p-measure}}
    \quad
    \subfloat[]{\includegraphics[width=0.45\textwidth]{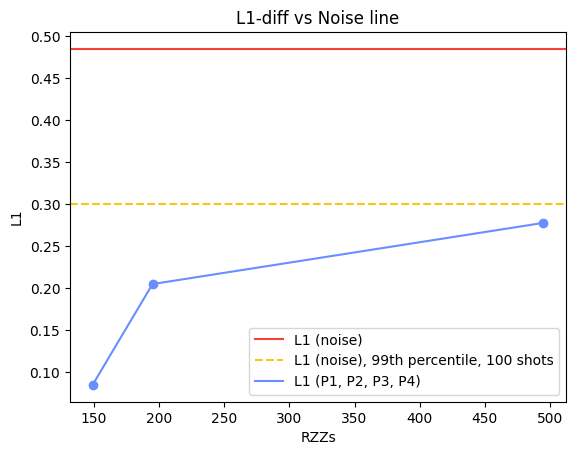}\label{fig:quantinuum-l1-diff}}
    \caption{(a) Statistics for the 4 most significant qubits for each of the 3 Quantinuum H1 experiments. The ideal measurement is the probability of measuring that qubit in the 0 state. (b) $L_1(P_{\{1..4\}})$ vs number of CNOTs. The red line is the expected value for $L_1(P_{\{1..4\}})$ if we had the maximally mixed quantum state. The yellow line is the 99th percentile of that statistic assuming we had 100 shots.}
    \label{fig:quantinuum-stats}
\end{figure}

We plot the probabilities $P_1, P_2, P_3, P_4$ for each of the circuit sizes in Figure~\ref{fig:quantinuum-p-measure}. The ideal simulator values $P^{*}_{i}$ do not differ much at all between the different circuits, allowing us to directly compare the results across each of the runs. Not surprisingly, the deeper circuits are still affected by noise, leading to an overall reduction in the accuracy of these coarse grained statistics. 

To quantify this relationship we can look at $L_1$ difference between the quantum experiments and ideal simulator values over a defined subset of qubits: 

\begin{equation}
L_1(P_{S}) = \sum_{i \in S} |P_i - P^{*}_i|
\label{eq:l1}
\end{equation}

We plot $L_1(P_{\{1..4\}})$ depending on the number of RZZ gates used in Figure~\ref{fig:quantinuum-l1-diff}. The general trend shows that as the circuits get deeper, the $L_1$ gets larger, indicating a larger discrepancy between the ideal simulation. However, even for the circuit with 495 RZZ gates, we still obtain expectation values far from what maximally mixed individual qubits would yield.

In Figure~\ref{fig:quantinuum-p-measure}, we benchmark our performance on this coarse-grained statistic by comparing to statistical fluctuations of a maximally mixed state. In the maximally mixed state, all $P_i$ are 0.5, leading to a $L_1$ difference of $0.48$. However, due to limited number of shots, one could measure $P_i$ that deviate from the true expectation value. By  simulating many such experiments with finite shots, we find that the 99-th percentile value of this $L_1$ is at $\sim 0.30$, indicating that even our deepest circuits are far from producing random bits. Note this analysis holds even if the noise is depolarizing on single qubits as opposed to the global state. 

\begin{figure}[ht]
\end{figure}

\subsection{IBM}
Here we present our image loading circuits on IBM hardware, namely the Algiers and Brisbane chips featuring 27 and 127 qubits, respectively. In Section~\ref{sssec:ibm-naive}, we discuss our experiments loading a 21 qubit state to encode the full image from the Honda Scenes dataset, which we refer to as the naive loading. In Section~\ref{sssec:ibm-bae}, we expand the circuit to 72 qubits by first down-sampling the image and segmenting it into 6 blocks. Again, since we have relatively limited shots compared to the granularity of the loaded image, we will focus on coarse grained statistics. For the full naive loading of the image, we probe the joint distribution over the first 3 most significant bits.

In our BAE strategy, we find that many of the marginals for the single qubit $Z$ measurements look uniform. Hence, to better distinguish these experimental outcomes from noise, we employ a strategy to learn the correct basis for maximizing the deviation from $0.5$. Similar to Section~\ref{ssec:quantinuum}, we then benchmark the performance of our methods as a function of increasing circuit depth, using two-qubit gates as the primary axis.

\subsubsection{Naive Loading}\label{sssec:ibm-naive}
\begin{figure}[ht]
    \centering
    \includegraphics[width=0.55\textwidth]{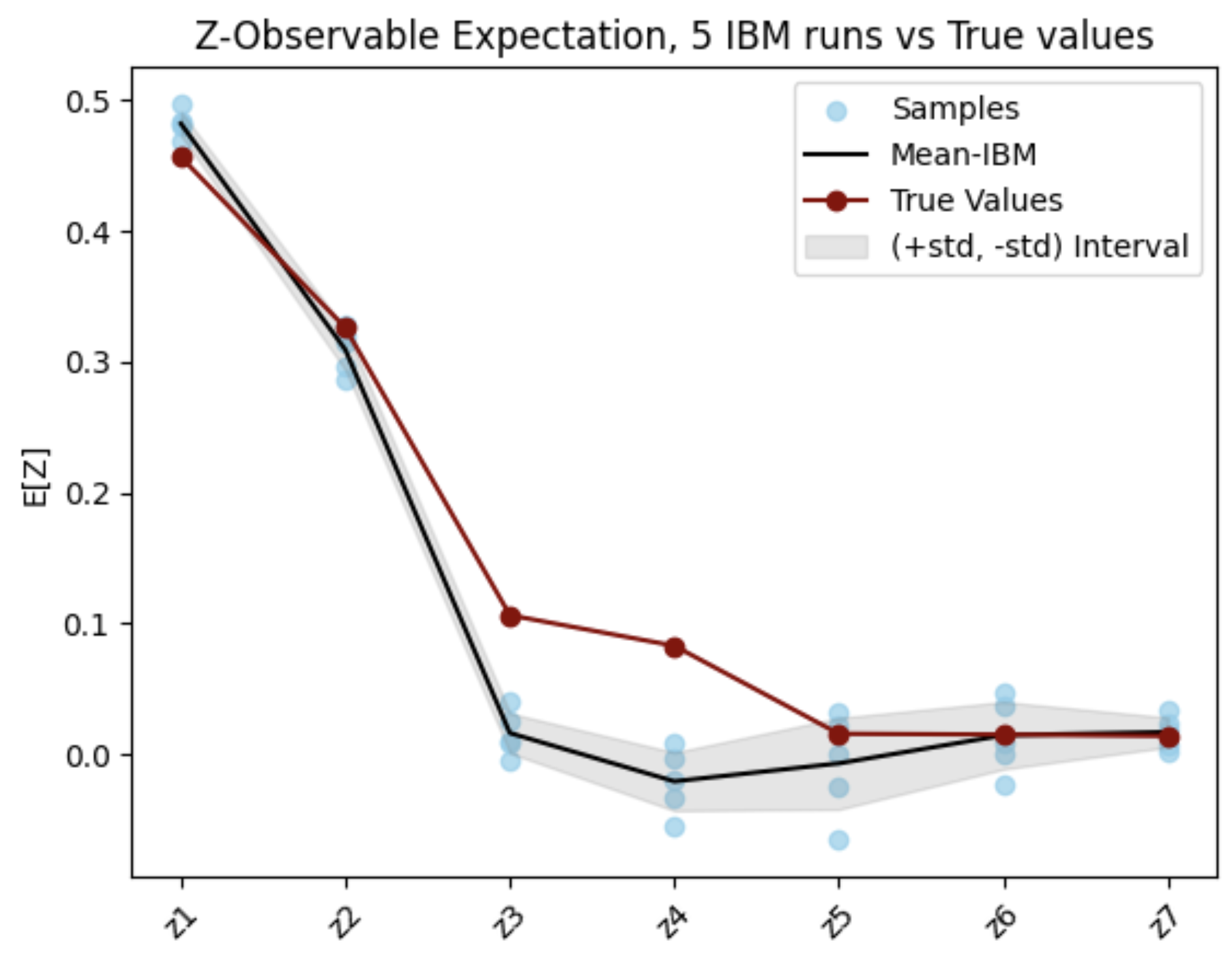}
    \caption{The red line depicts the actual values obtained from Pauli Z measurements on the encoded image. The blue dots illustrate the outcomes of various experimental runs on IBM's 27-qubit computer, including the mean and variance of these results. }
    \label{fig:ibm-measure}
\end{figure}

Using our simulators, we construct and train a hierarchical ansatz featuring the available RY, RZ, and CNOT gates on 21 qubits to encode a single $1024 \times 2048 $ image. The circuit contains a total of 229 gates, 80 two-qubit gates and 149 single-qubit gates. This was then deployed on IBM-Algiers with the single qubit $Z$ expectation values plotted against their simulated ideal values plotted in Figure~\ref{fig:ibm-measure}.

\begin{figure}[ht]
    \centering
    \subfloat[]{\includegraphics[width=0.45\textwidth]{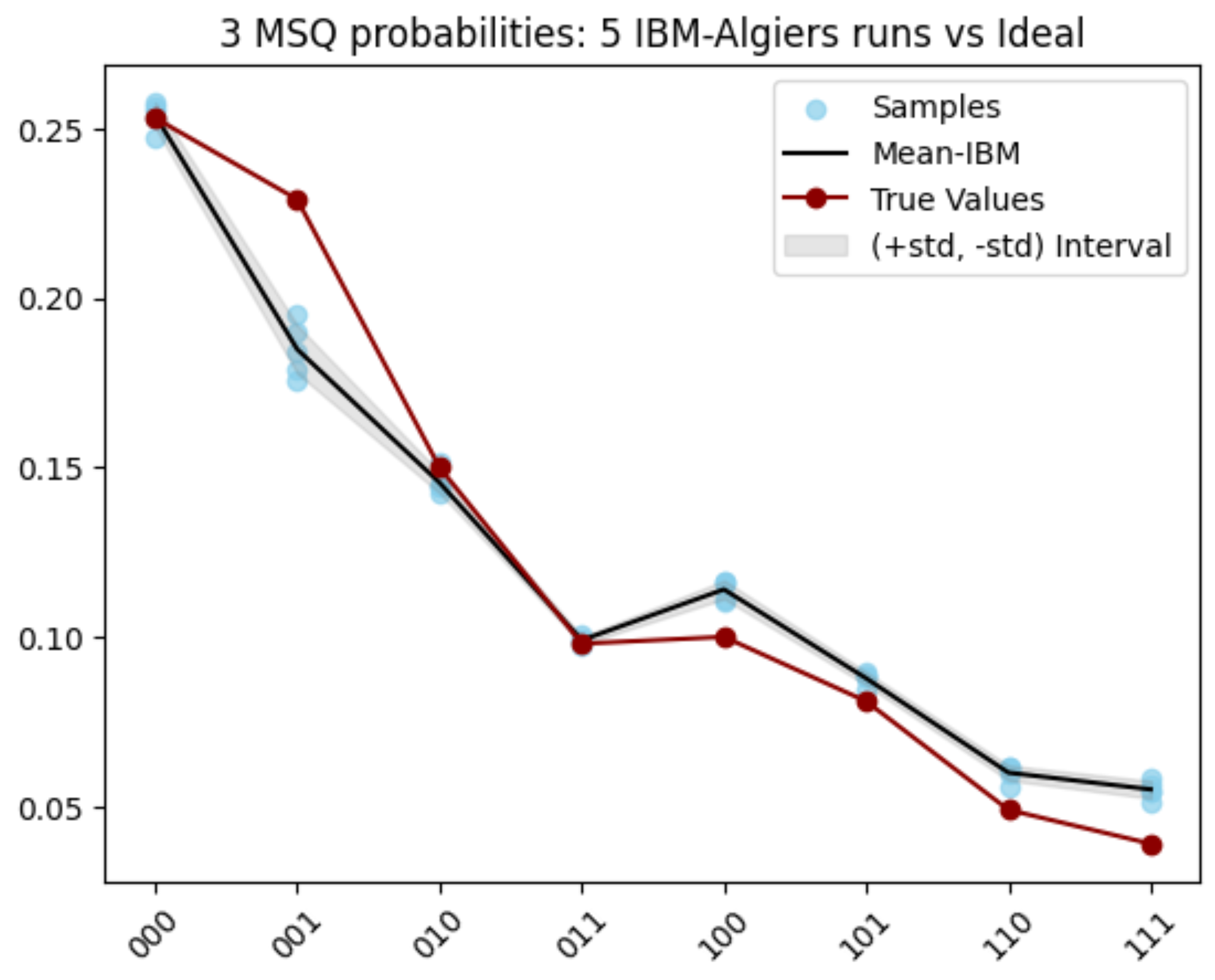}
    \label{fig:MSQ_3_ibm_algiers}}
    \quad
    \subfloat[]{\includegraphics[width=0.45\textwidth]{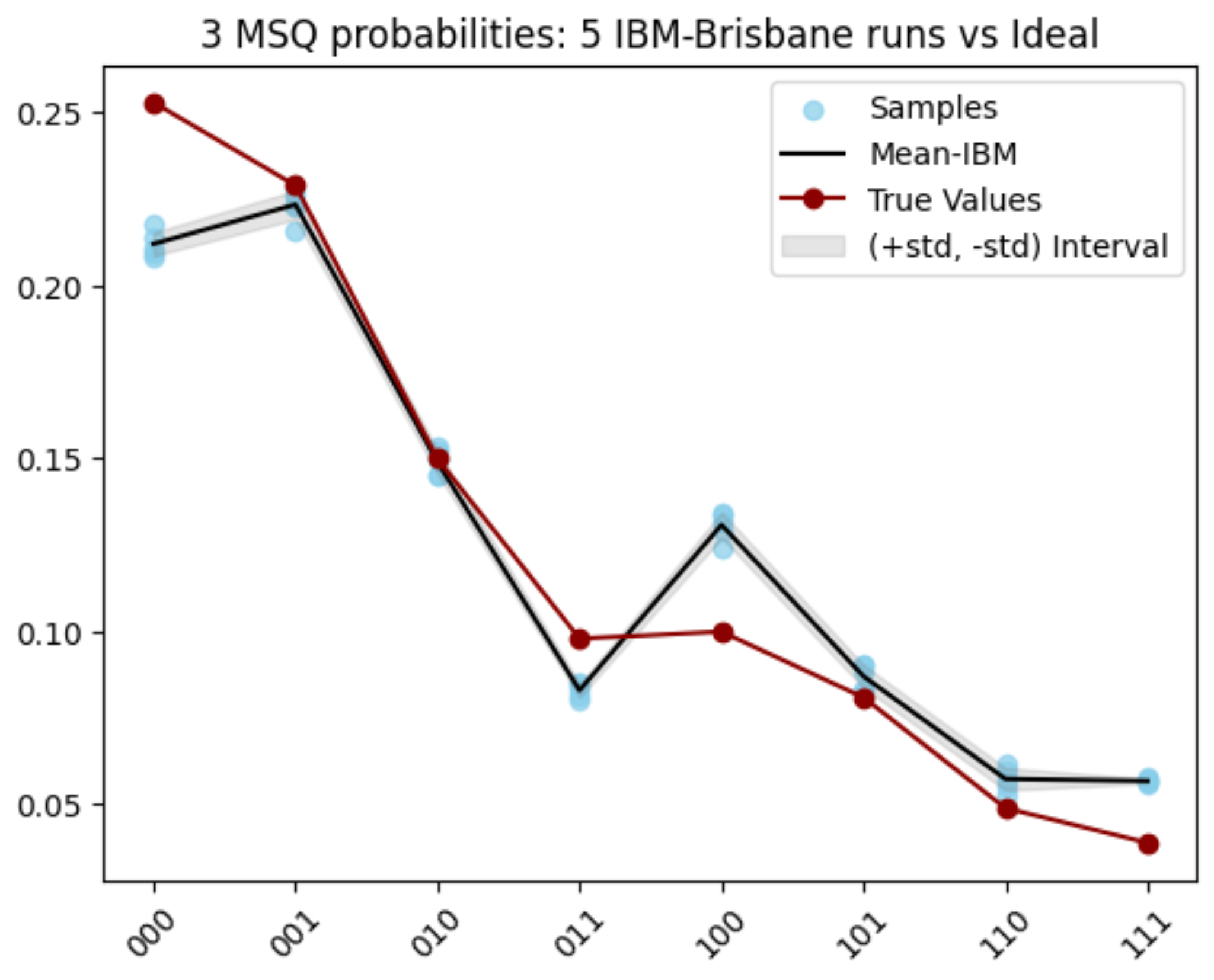}\label{fig:MSQ_3_ibm_brisbane}}
    \caption{The red line depicts the actual values of bitstring probabilities for the encoded image. The blue dots illustrate the outcomes of various experimental runs on IBM's 27-qubit (a) and 127-qubit (b) computers, including the mean and variance of these results. }
\end{figure}

With access to more shots, we were able to analyze our distribution at a slightly more in-depth level. In particular, we studied the bitstring measurement outcomes for the most significant bits, as opposed to single qubit marginals. 
In Figures \ref{fig:MSQ_3_ibm_algiers} and \ref{fig:MSQ_3_ibm_brisbane} we showcase the comparison between the true probabilities of the most significant bitstrings and the experimental results for the 27-qubit IBM-Algiers and 127 qubit IBM-Brisbane. Notably, the 27-qubit machine provides impressively accurate results, largely attributed to the shallow and hardware-friendly design of our quantum circuit ansatz. Because these are the most significant bits, one can think of this distribution as providing a snapshot of a coarse-grained image. 

\subsubsection{BAE Loading}\label{sssec:ibm-bae}
\begin{figure}
    \centering
    \includegraphics[width=0.96\linewidth]{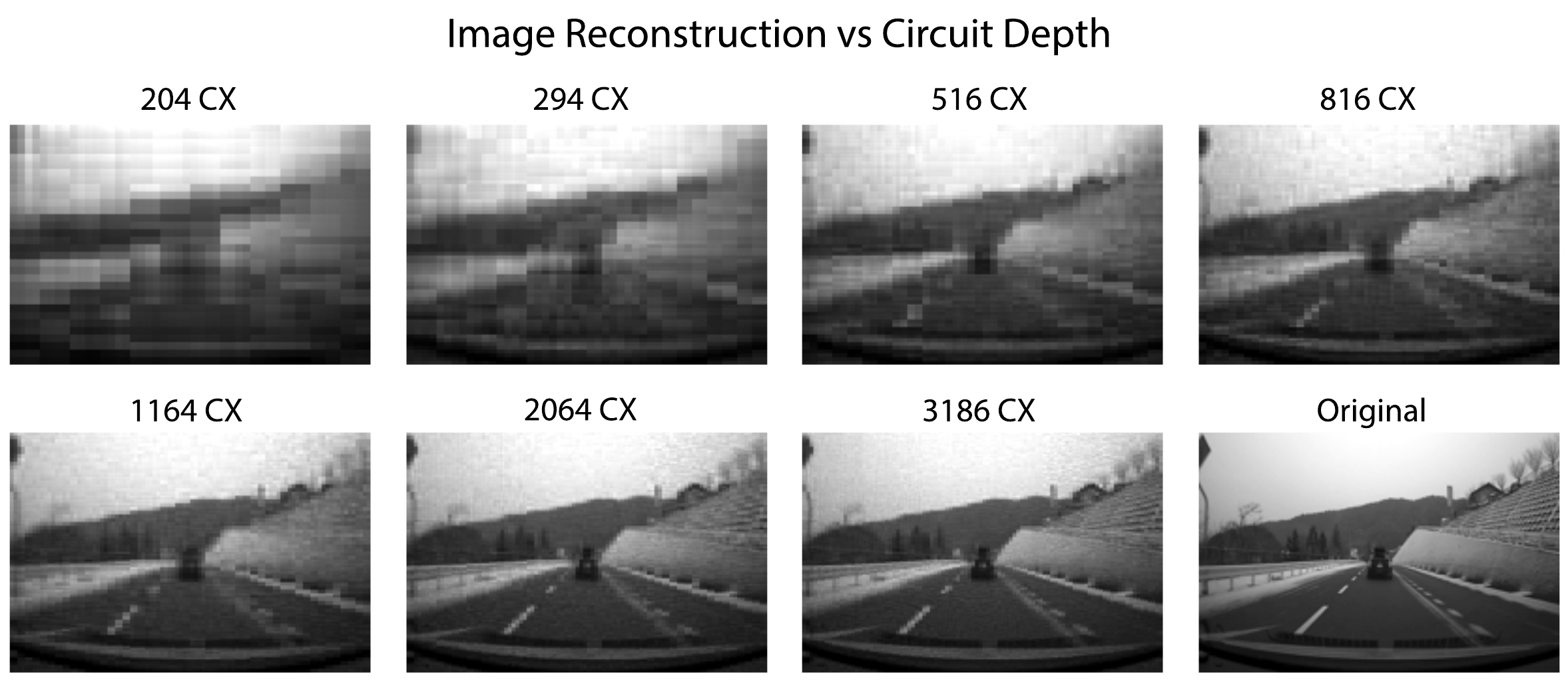}
    \caption{Loaded images for different number of CNOTs using BAE. In the smallest circuits, one can see rough shapes but no details in the cars or road. Around 1000 CX gates, lanes and a license plate start to become visible.}
    \label{fig:ibm-loaded-images}
\end{figure}
\begin{figure}
    \centering
    \includegraphics[width=0.96\linewidth]{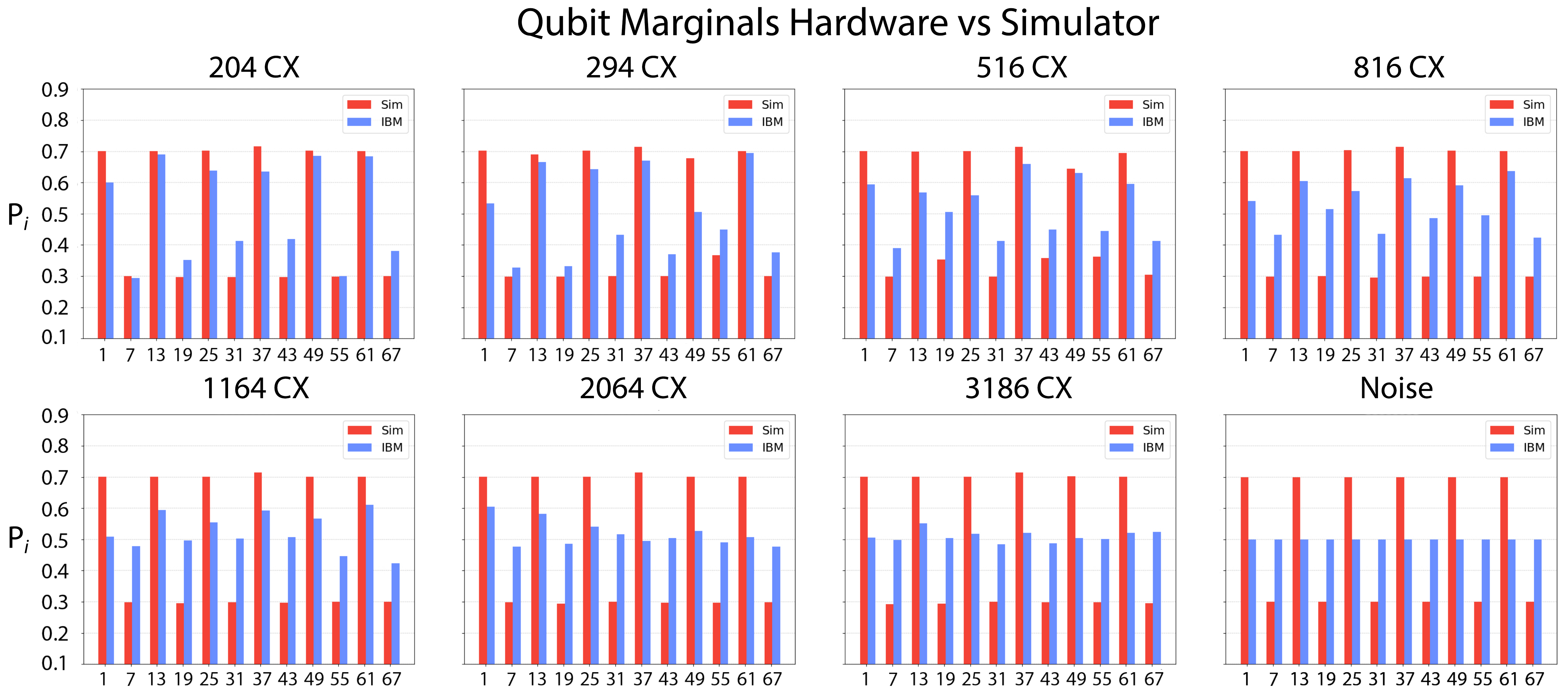}
    \caption{Comparison of statistics for IBM experiments with varying number of CNOTs. Red bars denote the probability of measuring a qubit in the 0 state in simulation and blue bars are experimental values. The final reference plot shows expected values of a completely noisy experiment. As CNOT count increases to around 3000, these single qubit measurements start to look maximally mixed.}
    \label{fig:ibm-comparison}
\end{figure}

\begin{figure}[h]
    \centering
    \includegraphics[width=0.55\textwidth]{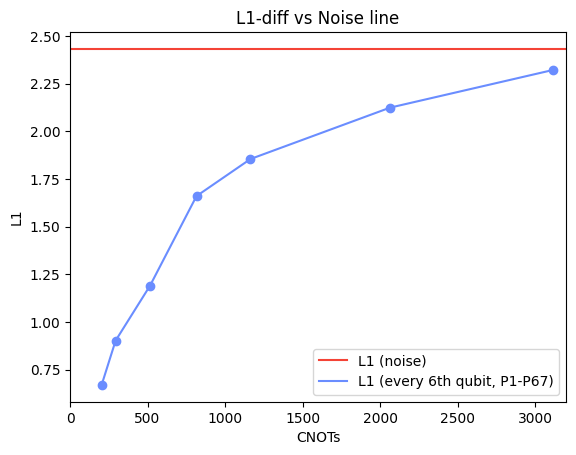}
    \caption{The blue line indicates the $L_1$ distance between the simulated and experimental histograms. The red line is the expected value for $L_1$ of a completely noisy experiment. Here we are not plotting the yellow 99-th percentile line for the noisy simulation since we have 4000 shots and the variance of the results is very small, making the 99th percentile line be very close to the red line. } 
    \label{fig:ibm-l1-diff}
\end{figure}

For the block-amplitude encoded version of the experiment, we chose to utilise the 127-qubit IBM Brisbane system, as it seems to provide an optimal trade-off between available qubits and the gate error rate. By using the different blocks, we maximize the usage of qubits while keeping the overall circuit depth low, since the blocks can be loaded independently. We first rescale the original image to $128\times192$ pixels, which is then divided into $6$ blocks of $64\times64$ pixels. This means that to load a single block $12$ qubits are needed, amounting to a total of $72$ qubits. After a local simulated training is performed for each of the blocks separately, the trained subcircuits are combined into a single large circuit, which is the quantum representation of the original image.

In order to evaluate the performance, we use the expectation values of the Z observable on the first and the seventh qubit of each of the subcircuits. These correspond to the most significant qubits in the vertical and horizontal directions, respectively. While trying to estimate these values locally with a simulator, we noticed that for some qubits they are very close to $0$, making them indistinguishable from the hardware depolarizing noise. Instead of running the original circuits we employed a strategy which would allow us to distinguish these values from noise, thus comparing the hardware and the simulated results at a slightly increased cost of 2-qubit gates.

This strategy involves ``learning'' an observable which has a non-trivial expectation value. After adding lightweight parametrized and entangling layers at the end of the subcircuits, another round of training is performed for each of them. The cost function is defined to be the mean squared error between the current expectation value of the Z observable and one of the possible extremes ($-1$ or $1$, depending on the qubit). The cost is optimized only over the newly added parameters, separately for each of the qubits. The training is cut off once the simulated expectation values deviate from the maximally mixed state by $0.4$.

We ran $7$ experiments, with an increasing number of CNOT gates, starting with a smallest of $204$ CNOTs (across all blocks) and going up to $3114$ CNOTs. The loaded images for these setups are presented on Figure \ref{fig:ibm-loaded-images}. For each experiment, we calculate the $P_i$ based on the expectation values obtained for each qubit on 4000 measurement shots. We present the results as histograms in Figure \ref{fig:ibm-comparison}.

We can see how with the increasing number of CNOT gates the experimental values for $P_i$ are getting closer to $0.5$ indicating the dominance of noise in the larger experiments. To better quantify this distance from a completely noisy experiment, we once again calculate the $L_1$ distance between the simulated and experimental expectation values and compare it with the baseline $L_1$ distance between the simulated and completely noisy histogram (with all the probabilities equal to $0.5$) based on equation \ref{eq:l1}, with $S = \{6k + 1, k \in \{0...11\} \}$. The results are shown in Figure~\ref{fig:ibm-l1-diff}. Indeed, this figure shows that with the increasing number of 2-qubit gates, we are getting closer to the baseline value of the noisy experiment. Nevertheless, this also shows that some signal can be extracted even at 2000 CNOTs.

\section{Discussion}\label{sec:discussion}
In this work, we have successfully extended our hierarchical circuit ansatz to load classical image data into quantum registers. 
Our methods allow us to train and implement image loading on 21 qubit circuits on existing quantum hardware, compared to the 10 qubit circuits for MNIST. 
Utilizing BAE, we can load images with more precision at the expense of larger total qubit counts. As long as each block remains relatively small, however, even a circuit training procedure on 72 qubits can be implemented.

Part of our approach differs in that we train on the resulting classical distribution of the quantum state. Namely, we do not place particular emphasis on the phases and the learned amplitudes are related to the square roots of the data. One could modify this by including the measurement distribution in a different basis as part of the training. In either case, amplitude encoding can be thought of as a feature map from the original data space to the Hilbert space of the qubits. So far, the strengths and weaknesses of different feature maps in the quantum setting remain under explored.

Data loading is a key bottleneck in the pipeline of quantum machine learning, especially when trying to compare to classical ML performance. Our methods provide a resource efficient way to train and deploy circuits. Though there is an upfront cost to training the circuits, they can be reused easily once training has been done. Having efficient representations of the classical data is very important for building quantum classifiers as the depth of quantum circuits is still limited by noise.

One may also consider using similar techniques presented in this paper to come up with generative models that encode multiple images at once and make it possible to generate new images or data points every time the quantum state is sampled.

Ultimately, we believe the hierarchical ansatz is an effective strategy for avoiding barren plateaus as one can always start in the small qubit regime before growing to larger circuit sizes. Here, we train numerically on the KL divergence, but in future work, one may also consider the role of other losses in the training process.

\textit{Author Contributions:} H. G. contributed to the concept, ideation, algorithm design, and paper editing. H. K. handled data preparation, data processing, and the implementation of the classical CNN model. T. S. was responsible for IBM deployment, circuit engineering, and evaluation. P. S. contributed to the concept, ideation, data selection, and paper editing. V. P. S. worked on algorithm design, quantum circuit implementation, and paper editing. R. H. T. focused on engineering work on CPU/GPU systems and evaluation. Finally, H. T. led the concept and algorithm implementation, served as the engineering lead, and managed the Quantinuum deployment.

\bibliographystyle{alpha}
\bibliography{refs_gibbs}

\appendix
\section{Software and Hardware details}\label{sec:software_hardware_details}
Below is the tech stack used for the experiments described in this paper:

\textbf{Pennylane for quantum circuits and circuit differentiation parts.} We use the adjoint differentiation techniques implemented by Pennylane when training our circuits to optimize the parameters of the ansatz.

\textbf{Nvidia H100 GPUs for gpu simulation}. For large circuit simulations of 20+ qubits the gpu simulators and Nvidia's cuQuantum library were used to achieve high performance and shorten the runtimes. To report specific numbers, we have been able to train 720 parameter, 20 qubit circuits for 1000 iterations in 300 seconds.

\textbf{IBM's 27 qubit (Algiers) and 127 qubit (Brisbane) quantum computers.} These are IBM's latest quantum chips that are publicly available and boast $>99\%$ median 2-qubit gate fidelity as well as $>99.9\%$ median 1-qubit gate fidelity.

\textbf{Quantinuum's H1 quantum computer.} This is one of the latest publicly available quantum chips from Quantintuum. H1 has 20 fully connected qubits with $99.998\%$ 1-qubit gate fidelities and $99.9\%$ 2-qubit gate fidelity. Quantinuum uses trapped ions for its qubits which are in general slower than superconducting qubits but instead have better qualities and connectivity.

\textbf{BlueQubit SDK for Hierarchical Loading and overall orchestration.} We used BlueQubit's SDK that has the implementation of hierarchical learning for various connectivities and ansatzes. Also the BlueQubit SDK "stitches" together all the libraries and devices mentioned above to make the experimentation needed for the results presented in this paper more seamless and efficient.

\end{document}